\RequirePackage{fix-cm}
\documentclass[twocolumn]{svjour3}          
\smartqed  
\usepackage{natbib}
\usepackage{graphicx}
\usepackage{enumerate}

\usepackage[table,xcdraw]{xcolor}
\usepackage{multirow}
\usepackage{rotating}
\usepackage{pifont}
\usepackage{notes}
\usepackage{listings}    
\usepackage{hyperref}
\usepackage{comment}
\usepackage{makecell}
\usepackage{appendix}

\usepackage[utf8]{inputenc}

\begin{document}

\title{A Comparative Survey of Recent Natural Language Interfaces for Databases}

\author{Katrin Affolter$^1$        \and
        Kurt Stockinger$^1$         \and
        Abraham Bernstein$^2$ 
}


\institute{$^1$Zurich University of Applied Sciences, Switzerland       \\
           $^2$University of Zurich, Switzerland
}

\date{Received: date / Accepted: date}

\maketitle

\begin{abstract}
Over the last few years natural language interfaces (NLI) for databases have gained significant traction both in academia and industry. These systems use very different approaches as described in recent survey papers.  However, these systems have not been systematically compared against a set of benchmark questions in order to rigorously evaluate their functionalities and expressive power.

In this paper, we give an overview over 24 recently developed NLIs for databases. Each of the systems is evaluated using a curated list of ten sample questions to show their strengths and weaknesses. We categorize the NLIs into four groups based on the methodology they are using: keyword-, pattern-, parsing-, and grammar-based NLI. Overall, we learned that keyword-based systems are enough to answer simple questions. To solve more complex questions involving subqueries, the system needs to apply some sort of parsing to identify structural dependencies. Grammar-based systems are overall the most powerful ones, but are highly dependent on their manually designed rules. In addition to providing a systematic analysis of the major systems, we derive lessons learned that are vital for designing NLIs that can answer a wide range of user questions.

\keywords{Natural language interfaces \and Query processing \and Survey}
\end{abstract}

\section{Introduction}
\label{sec:introduction}

Living in a digital age, the global amount of data generated is increasing rapidly. A good part of this data is (semi-)structured and stored in some kind of database, which is usually accessed using query languages such as SQL or SPARQL. Structured query languages, however, are difficult to understand for non-experts. Even though SQL was initially developed to be used by business people, reality shows that even technically-skilled users often experience problems putting together correct queries \citep{Bowen2004}, because the user is required to know the exact schema of the databases, the roles of various entities in the query and the precise join paths to be followed. Non-technical (or casual) users are usually overwhelmed by the technical hurdles of formal query languages.

One often-mentioned approach to facilitate database querying even for casual users is the use of \textit{natural language interfaces} (NLI) for databases. These allow users to access information stored in databases by typing questions expressed in natural language \citep{hendrix1978developing}. Some NLIs restrict the use of the natural language to a sub-language of the domain or to a natural language restricted (and sometimes controlled) by grammatical constraints. For example, SODA \citep{SODA2012}, is based on English keywords, ATHENA \citep{Athena2016} handles full sentences in English, and Ginseng \citep{Ginseng2005} strictly guides the user to a correct full sentence in English. If users want to find all movies with the actor Brad Pitt, for instance, the input question in the different systems could be as follows:

\begin{description}
	\item [SODA:] \texttt{movie Brad Pitt}
	\item [ATHENA:] \texttt{Show me all movies with the actor\\Brad Pitt.}
	\item [Ginseng:] \texttt{What are the movies with the actor\\Brad Pitt?}
\end{description}

\noindent Depending on the database schema, the corresponding SQL statement might be as follows:

\lstset{language=SQL,basicstyle=\ttfamily,} 
\begin{lstlisting}
SELECT m.title FROM Movie m 
  JOIN Starring s ON s.movieId = m.id 
  JOIN Actor a ON a.actorId = s.actorId 
  JOIN Person p ON p.id = a.actorId 
 WHERE p.FirstName = "Brad" 
   AND p.LastName = "Pitt"
\end{lstlisting}

\noindent As we can see in this example, the users of NLIs do not have to know the underlying structure or query language to formulate the question in English. 

Critiques of NLIs often highlight that natural language is claimed to be too verbose and too ambiguous. If it were possible to identify the different types of linguistic problems, then the system could support the user better, for example, with a clarification dialog. This would not only help the NLI to translate the natural language question into a formal query language, but would also assist the users in formulating correct queries. For example, if the users ask a question like `\textit{all movies from Columbia Pictures or 20th Century Fox}', the system should ask for clarification of the ambiguous token `\textit{Columbia}' which can be either a location or a movie company (as part of the bi-gram `\textit{Columbia Pictures}'). If the users choose movie company, then the system could directly suggest that `\textit{20th Century Fox}' is also a movie company and not the concept for films from 1900 until 1999.

Additionally, natural language is not only ambiguous on word-level, but there can be also multiple interpretations of the meaning of a sentence. In the example input question `\textit{all cinemas in cities with parking lots}' the constraint `\textit{with parking lots}' can either refer to the cinemas or to the cities. Both interpretations are grammatically correct, but probably the intent is to find cinemas with a parking area. In some cases, like `\textit{all cinemas in cities with love seats}' there is only one interpretation valid on the data set: only cinemas can have a constraint on `\textit{love seats}', even if it would be grammatically correct for cities. Therefore, an NLI should first check if multiple interpretations exist. Next, it should verify if they can be applied on the dataset. If there are still multiple interpretations left, the users should have the possibility to choose the most relevant one.

The above mentioned examples are only some of the linguistic problems that NLIs must deal with. Furthermore, users are not perfect and tend to make mistakes. These range from spelling errors to the use of colloquial language, which includes syntactically ill-formed input. These examples do, however, highlight that precisely understanding the expressiveness of questions that an NLI is able to interpret is of paramount importance both for developers and users of NLIs. To address this need, this paper makes the following \textbf{contributions}:

\begin{itemize}
\item We provide an overview of recent NLI systems comparing and analyzing them based on their expressive power.
\item Existing papers often use different data sets and evaluation metrics based on precision and recall, while others perform user studies to evaluate the system (see Section 5 for details). Given this heterogeneity of evaluations, it is very hard to directly compare these systems. Hence, we  propose a set of sample questions of increasing complexity as well as an associated domain model
aiming at testing the expressive power of NLIs.
\item The paper serves as a guide for researchers and practitioners, who want to give natural language access to their databases.
\end{itemize}

To help the reader understand the differences between current NLIs, in Section \ref{sec:sample_world} we first describe a sample world, which we use as an example running consistently through the paper, in order to evaluate and benchmark the systems. This is different to previous surveys on NLIs \citep{Gautam2017a,Li2017,Mishra2016,Shafique2014,Nihalani20011}, where the systems are only summarized or categorized. Based on our sample world, we introduce ten representative sample questions of increasing expressive complexity and illustrate their representativeness using question answering corpora. These questions are used to compare the NLIs and highlight the various difficulties the NLIs face. In Section \ref{sec:nlp_technologies} we give a brief overview of the most important technology for natural language processing and discuss the limitations of our evaluation in Section \ref{sec:limitations}. The sample world and the questions are then used in Section \ref{sec:systems} to explain the process of translating from natural language questions to a formal query language for each NLI surveyed. We provide a systematic analysis of the major systems used in academia (Section 6.1) as well as three major commercial systems (Section 6.2). In Section 7 we discuss newer developments of NLIs based on machine learning. Finally, we derive lessons learned that
are vital for designing NLIs that can answer a wide range of user questions (Section 8).

Note that the detailed evaluation of the systems in Section 6 omits the approaches based on machine learning. The rationale for this is twofold. First, the functionality of the machine learning based approaches is heavily dependent on the training data. Most papers present a system trained with some dataset and then show the capabilities of the system in the context of that training. We found no exploration of the general capabilities or robustness of a given approach when varying the input data. Hence, but is difficult to say if these systems could cover all requirements when given suitable training data or not. Second, little is known how domain dependent those systems are on the training data. Usually they require a lot of training examples making the comparison to the other systems—that mainly require some meta-data—difficult. We, hence, concluded that the categorization of the capability of these systems is an open problem that needs to be addressed in its own survey.

\section{Foundation: A Sample World}
\label{sec:sample_world}

In this section, we present a small sample world that serves as the basis for analyzing different NLIs. We first introduce the database schema (Section \ref{subsec:database_ontology}) and afterwards discuss ten input questions of increasing complexity that pose different challenges to NLIs (Section \ref{subsec:input_questions}). Finally, we will  perform an analysis of different question-answering corpora to better understand what types of questions real users pose and how we can map them to our ten input questions. Our analysis in Section \ref{subsec:question_analysis} indicates that our ten sample questions represent a large range of questions typically queried in question answering systems.

\subsection{Database Ontology}
\label{subsec:database_ontology}

\begin{figure}
	\includegraphics[width=\columnwidth]{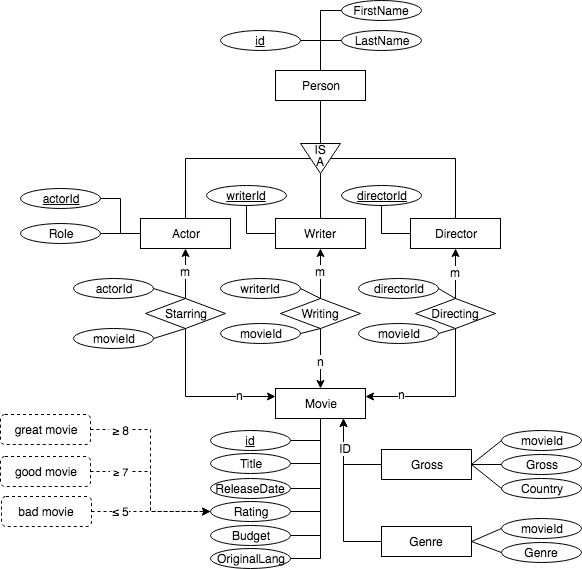}
	\caption{Ontology of the sample world `movie database'.}
	\label{fig:sample_world}
\end{figure}

The sample world is a small movie database inspired by IMDB\footnote{\url{https://www.imdb.com/}} and extended with hierarchical relationships and semantic concepts. Figure \ref{fig:sample_world} visualizes the ontology of the sample world, which consists of movies, persons, and their relationships employing an entity-relation diagram. A person can be an actor, a writer, and/or a director. For movies, information about starring actors, writers and directors is stored. Each movie has a title, a release date, a rating (float), a budget (bigint) and the original language. It can also have multiple genres and multiple gross profits for different countries. Furthermore, there are three concepts --- \textit{bad}, \textit{good}, and \textit{great} movie --- defined, all depending on the rating of the movies. Elements with solid lines correspond to the schema of the underlying database. The entries in the database will be called \textbf{base data}. The concepts and the schema are the \textbf{meta data}. To improve performance, many NLIs implement inverted indices over literal values such as strings (both from the base and meta data). Elements with dotted lines correspond to (possibly derived) concepts, which are only defined in the ontology. 

\begin{figure}
	\includegraphics[width=\columnwidth]{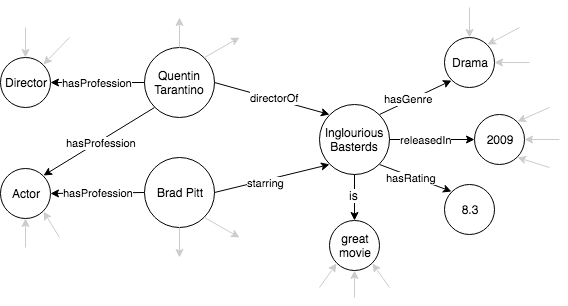}
	\caption{Part of the knowledge graph representing the sample world `movie database'.}
	\label{fig:knowledge_graph}
\end{figure}

Another possible representation of the data would be as a knowledge graph. The core concepts of a knowledge graph are the entities and their relationships. Figure \ref{fig:knowledge_graph} visualizes a part of the sample world as a knowledge graph. Concepts can be directly included in the knowledge graph. For example, the movie `\textit{Inglourious Basterds}' has a link to the concept `\textit{great movie}'.  

\subsection{Input Questions}
\label{subsec:input_questions}

To explain the different NLIs and highlight their\linebreak strengths and weaknesses, input questions are necessary. NLIs are designed to help non-tech users to formulate formal queries using natural language. The design of our questions is inspired by Turing Award Winner Jim Gray who designed 20 challenging questions of increasing complexity that were used as the yard stick for evaluating large-scale astrophysics databases (\cite{szalay2002sdss}). This methodology of designing questions of increasing complexity also served as the basis for the evaluation of SODA (\cite{SODA2012}) and the 10 sample questions of this paper.

Therefore, we designed nine questions based on the operators of SQL and SPARQL: \textbf{J}oins, \textbf{F}ilters (\textbf{s}tring, \textbf{r}ange, \textbf{d}ate or \textbf{n}egation), \textbf{A}ggregations, \textbf{O}rdering, \textbf{U}nion and \textbf{S}ubqueries.

Furthermore, we added a question which is based on a \textbf{c}oncept (e.g., `\textit{great movie}'). \textbf{C}oncepts are not part of SQL and SPARQL, but a common addition of NLIs. Table \ref{tab:input_questions} shows those ten full sentence questions in English, which can be applied on the sample world (ordered roughly by difficulty).

The queries were designed in such a way that they cover a wide range of SQL-functionality (technical challenge) as well as linguistic variations (semantic challenge). We will now analyze each of these ten queries in more detail and describe the major challenges for the underlying system to solve them.

\begin{table}
\centering
\caption{Ten sample input questions based on SQL/SPARQL operators that are answerable on the sample world. (\textbf{J}oin; \textbf{F}ilter (\textbf{s}tring, \textbf{r}ange, \textbf{d}ate or \textbf{n}egation);  \textbf{A}ggregation; \textbf{O}rdering; \textbf{U}nion; \textbf{S}ubquery; \textbf{C}oncept)}
\label{tab:input_questions}
\begin{tabular}{lll}
\# & natural language question & challenges\\
\hline
\texttt{Q1} & \begin{tabular}[t]{@{}l@{}}Who is the director of ``Inglourious \\Basterds''?\end{tabular} & \texttt{J, F(s)} \\
\texttt{Q2} & All movies with a rating higher than 9. & \texttt{J, F(r)} \\
\texttt{Q3} & \begin{tabular}[t]{@{}l@{}}All movies starring Brad Pitt from 2000\\ until 2010.\end{tabular} & \texttt{J, F(d)} \\
\texttt{Q4} & Which movie has grossed most? & \texttt{J, O} \\
\texttt{Q5} & Show me all drama and comedy movies. & \texttt{J, U} \\
\texttt{Q6} & List all great movies. & \texttt{C} \\
\texttt{Q7} & What was the best movie of each genre? & \texttt{J, A} \\
\texttt{Q8} & List all non-Japanese horror movies. & \texttt{J, F(n)} \\
\texttt{Q9} & \begin{tabular}[t]{@{}l@{}}All movies with rating higher than the \\rating of ``Sin City''.\end{tabular} & \texttt{J, S} \\
\texttt{Q10} & \begin{tabular}[t]{@{}l@{}}All movies with the same genres as ``Sin \\City''.\end{tabular} & \texttt{J,} 2x\texttt{S}
\end{tabular}
\end{table}

The first question (\texttt{Q1}) is a join over different tables (\texttt{Person}, \texttt{Director}, \texttt{Directing} and \texttt{Movie}) with an ISA-relationship between the tables \texttt{Person} and \texttt{Di}-\texttt{rector}. Moreover, the query has a filter on the attribute \texttt{Movie.Title}, which has to be equal to `\textit{Inglourious Basterds}'. Therefore, the system faces three different challenges: (a) identify the bridge table \texttt{Directing} to link the tables \texttt{Director} and \texttt{Movie}, (b) identify the hierarchical structure (ISA-relationship) between \texttt{Director} and \texttt{Person} and (c) identify `\textit{Inglourious Basterds}' as a filter phrase for \texttt{Movie.Title}.

The second question (\texttt{Q2}) is based on a single table (\texttt{Movie}) with a range filter. The challenge for the NLIs is to translate `\textit{higher than}' into a comparison operator `\textit{greater than}'.

The third question (\texttt{Q3}) is a join over four tables (\texttt{Person}, \texttt{Actor}, \texttt{Starring} and \texttt{Movie}) and includes two filters: (a) a filter on the attribute \texttt{Person.FirstName} and \texttt{Person.LastName} and (b) a two-sided date range filter on the attribute \texttt{Movie.ReleaseDate}. The challenge in this query (compared to the previous ones) is the date range filter. The system needs to detect that `\textit{from 2000 until 2010}' refers to a range filter and that the numbers need to be translated into the dates \textit{2000-01-01} and \textit{2010-12-31}.

The fourth question (\texttt{Q4}) is a join over two tables (\texttt{Movie} and \texttt{Gross}). In addition, an aggregation on the attribute \texttt{Gross.Gross} and grouping on the attribute \texttt{Movie.id} or ordering the result based on \texttt{Gross.Gross} is needed. For both approaches an aggregation to a single result (indicated by the keyword `\textit{most}') is requested.

The fifth question (\texttt{Q5}) is a join over two tables (\texttt{Movie} and \texttt{Genre}). The query can either be interpreted as `\textit{movies that have both genres}' (intersection) or `\textit{movie with at least one of those genres}' (union). The expected answer is based on the union interpretation, which can be solved with two filters that are concatenated with an OR on the attribute \texttt{Genre.Genre}.

The sixth question (\texttt{Q6}) needs the definition of concepts. In the sample world the concept `\textit{great movie}' is defined as a movie with a rating greater or equal 8. If the system is capable of concepts, then it needs to detect the concept and translate it accordingly to the definition.

The seventh question (\texttt{Q7}) is a join over two tables (\texttt{Movie} and \texttt{Genre}) with an aggregation. The challenges are to (a) identify the grouping by the attribute \texttt{Genre.Genre} and (b) translate the token `\textit{best}' to a maximum aggregation on the attribute \texttt{Movie.Rating}.

The eighth question (\texttt{Q8}) is a join over two tables (\texttt{Movie} and \texttt{Genre}) with a negation on the attribute \texttt{Movie.OriginalLang} and a filter on the attribute \linebreak \texttt{Genre.Genre}. The challenge in this question is to identify the negation `\textit{non-Japanese}'. Another possible input question with a negation over a larger movie data-base, would be `\textit{All actors without an Oscar}'. Here again, the challenge is to identify `\textit{without}' as a keyword for the negation.

The ninth question (\texttt{Q9}) is based on a single table (\texttt{Movie}) and includes a subquery. The challenge in this question is to divide it in two steps: first select the rating of the movie `\textit{Sin City}' and then use this SQL statement as a subquery to compare with the ranking of every other movie in the database. 

The tenth question (\texttt{Q10}) is a join over two tables (\texttt{Movie} and \texttt{Genre}). One possible solution would include two \texttt{not exist}: the first one verifies for each movie that there exist no other genres as the genres of `\textit{Sin City}'. The second one verifies for each movie that it has no genre, which `\textit{Sin City}' does not have. For example, the movie `\textit{Sin City}' has the genre `\textit{Thriller}', the movie `\textit{Mission: Impossible}' has the genres `\textit{Thriller}' and `\textit{Action}'. The first \texttt{not exist} will check if `\textit{Mission: Impossible}' has the genre `\textit{Thriller}' from `\textit{Sin City}' which is true. The second \texttt{not exist} checks if `\textit{Sin City}' has the genres `\textit{Thriller}' and `\textit{Action}' (from `\textit{Mission: Impossible}'), which is false.

\subsection{Question Analysis}
\label{subsec:question_analysis}

In this section, we perform an analysis comparing our ten sample input questions with the two well-known question-answering corpora Yahoo! QA Corpus L6\footnote{\url{https://webscope.sandbox.yahoo.com/catalog.php?datatype=l}}\linebreak (more than 4 million questions within a user community) and GeoData250 \citep{Mooney2001} (250 questions against a database). We also summarize the findings of \cite{Bonifati2017} and compare them to our input questions. The goal of the analysis is to better understand what types of questions users pose and how representative our sample input questions are (i.e.,~establish some external validity of their representativeness).

For the Yahoo! Corpus, we decided to only look into the labeled subset of movie questions since our sample world is based on movies. Out of these questions related to movies we extracted a random sample set of 100 questions. We used all the GeoData250 Questions. We labeled each of those 350 questions from both sources with \texttt{Q1} to \texttt{Q10} based on what poses the challenge in answering them. 

For example, the GeoData250 question `\textit{give me the cities in virginia?}' is labeled with \texttt{Q1}, because the challenge of this question is to identify the right filter (`\textit{virginia}'). The question `\textit{what is a good movie to go see this weekend?}' includes a time range and is therefore labeled as \texttt{Q6}. For the Yahoo! Corpus we also made some assumptions, for example, the question `\textit{what is your favorite tom hanks movie?}' is interpreted as `\textit{give me the best ranked tom hanks movie}' and labeled with \texttt{Q4}. Furthermore, if a question could have multiple labels, the label of the more difficult (higher number) question is chosen.  For example, the sample question `\textit{can anyone tell a good action movie to watch?}' is labeled with \texttt{Q6} because it requires handling of a concept (`\textit{good movie}') and not \texttt{Q1} because it uses a filter (`\textit{action movie}'). If the question cannot be labeled with one of the input questions, we label it with \texttt{x}\footnote{This label is only needed for the Yahoo! questions, where the topic is not about movies, even though they were labeled as 'movie' by Yahoo!}. For example, the question `\textit{i want to make a girl mine but she is more beautiful than me. what can i do now?}' has nothing to do with movies.

\begin{figure}
	\includegraphics[width=\columnwidth]{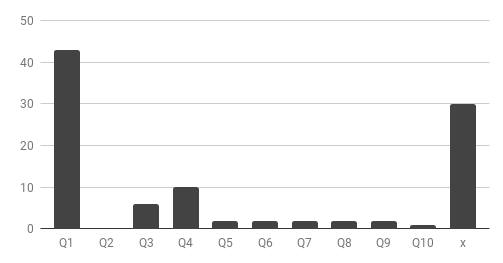}
	\caption{Mapping of 100 sample questions of Yahoo! QA Corpus L6 (movie) to our ten sample world questions.}
	\label{fig:chart_yahoo}
\end{figure}

As we can see in Figure \ref{fig:chart_yahoo}, more than 40\% of the Yahoo! questions are based on filtering only, i.e. corresponding to question \texttt{Q1}. For example, the question `\textit{what movie had ``wonderful world'' by sam cooke at the beginning?}' has filters for the song `\textit{wonderful world}' and a join on movie. About 30\% of the questions are labeled with \texttt{x} which are off-topic questions.  There are no questions labeled with \texttt{Q2}, this means that there are no questions with a numerical range. This can be explained by the composition of the corpus itself, which is a collection of questions from users to users. If users ask about the ranking of a movie, they ask something like `\textit{what is your favorite movie?}' and not something similar to \texttt{Q2}.

\begin{figure}
	\includegraphics[width=\columnwidth]{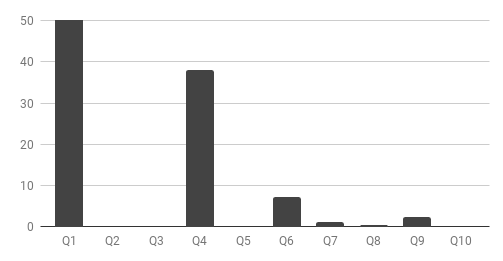}
	\caption{Mapping of the GeoQuery250 questions (in percentage) to our ten sample world questions.}
	\label{fig:chart_geodata}
\end{figure}

In Figure \ref{fig:chart_geodata}, we can see the distribution of question types for the GeoData250 corpus. About 88\% of the questions are labeled with \texttt{Q1} or \texttt{Q4}. There are three concepts (`\textit{population density}', `\textit{major city}' and `\textit{major river}') used in the corpus that occur in roughly 8\% of the questions, 7\% of which are labeled with \texttt{Q6}. There are no numerical range questions (\texttt{Q2}) and no date questions (\texttt{Q3}). The latter can be explained by the dataset not including any dates. There are also no unions (\texttt{Q5}) and no questions with multiple subqueries (\texttt{Q10}).

\cite{Bonifati2017} investigated a large corpus of query logs from different SPARQL endpoints. The query log files are from seven different data sources from various domains. One part of the analysis was done by counting the SPARQL keywords used in the queries. Over all domains, 88\% of the queries are \texttt{Select}-queries and 40\% have \texttt{Filter}. Furthermore, they found huge differences between different domains. For example, the use of \texttt{Filter} ranges from 61\% (LinkedGeoData) to 3\% (OpenBioMed) or less. This implies that the distribution of the usage for the question types are domain-dependent. Nevertheless, our ten sample question are fully covered in the query log analyzed by \cite{Bonifati2017}.

Our analysis indicates that our ten sample questions represent a large range of questions typically queried in question answering systems. Note that open-ended questions --- the ones we labeled as \texttt{x} in the Yahoo! Corpus --- are not covered by our analysis.

\section{Background: Natural Language Processing Technologies}
\label{sec:nlp_technologies}

In this section, we will discuss the most commonly used \textit{natural language processing} (NLP) technologies that are relevant for NLIs to databases. In particular we give a short overview on handling of stop words and synonyms, tokenization, part of speech tagging, stemming, lemmatization, and parsing.

\subsection{Stop Word}
\label{sec:stop_word}
The term \textit{stop word} refers to the most common words in a language. There is no clear definition or official list of stop words. Depending on the given purpose, any group of words can be chosen as stop words. For example, search engines mostly remove function words, like \textit{the}, \textit{is}, \textit{at} and others. Punctuation marks like dot, comma and semicolon are often included in the stop word list. For NLIs, stop words can contain invaluable information about the relationship between different tokens. For example, in the question `\textit{What was the best movie of each genre?}' (\texttt{Q7}) the stop words `\textit{of each}' imply an aggregation on the successive token `\textit{genre}'. On the other hand, stop words should not be used for lookups in the inverted indexes. In the question `\textit{Who is the director of ``Inglourious Basterds''?}' (\texttt{Q1}) the stop word `\textit{of}' would return a partial match for a lot of movie titles, which are not related to the movie `\textit{Inglourious Basterds}'. Therefore, the NLIs should identify stop words, but not remove them because they can be useful for certain computations.

\subsection{Synonymy}
\label{sec:synonymy}
The difficulty of synonymy is that a simple lookup or matching is not enough. For example, the question `\textit{All movies \texttt{starring} Brad Pitt from 2000 until 2010.}' (\texttt{Q3}) could also be phrased as `\textit{All movies \texttt{playing} Brad Pitt from 2000 until 2010.}'. The answer should be the same, but because in the sample world no element is named `\textit{playing}', a lookup would not find an answer. Therefore, it is necessary that the system takes synonyms into account. A possible solution is the use of a translation dictionary. Usually such a dictionary is based on \linebreak DBpedia \citep{DBpedia2012} and/or WordNet \citep{WordNet1995}.

\subsection{Tokenization}
\label{sec:tokenization}
Tokenization is used to split an input question into a list of tokens. It can be as simple as a separator on whitespace, but more often it is based on multiple rules (e.g.,~with regular expressions) or done with ML algorithms. Simple whitespace splitting tokenizers are often not good enough if the input question includes punctuation marks. For example, all input questions (\texttt{Q1-10}) end either with a question mark or a period. If the punctuation mark is not separated from the last word, the NLI would have to search for a match for the token `\textit{Basterds''?}' (\texttt{Q1}) instead of `\textit{Basterds}'. Without other processing, the NLI will not find any full matches. Depending on the task to solve, part of the tokenization process can be splitting on punctuation marks or deleting them. Either way, there are some scenarios to think about. For example, decimals should neither be split on the punctuation mark nor should they be removed.  Consider the following example `\textit{All movies with a rating higher than 7.5}' (similar to \texttt{Q2}). If you remove the dot between 7 and 5, the result would be completely different. Also other NLP technologies could be dependent on punctuation marks, for example, dependency trees.

\subsection{Part of Speech Tagging}
\label{sec:pos_tagging}
\begin{figure}
	\includegraphics[width=\columnwidth]{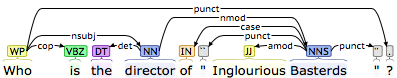}
    \caption[PoS tags and dependency tree for the sample question \texttt{Q1} (Stanford CoreNLP).]
    {PoS tags and dependency tree for the sample question \texttt{Q1} using Stanford CoreNLP.\\
    \textit{PoS tags}: \texttt{WP} = wh-pronoun; \texttt{VBZ} = verb, 3rd person singular present; \texttt{DT} = determiner; \texttt{NN} = noun, singular or mass; \texttt{JJ} = adjective; \texttt{NNS} = noun, plural \\
    \textit{Dependencies}: \texttt{punct} = punctuation; \texttt{nsubj} = nominal subject; \texttt{cop} = copula; \texttt{det} = determiner; \texttt{nmod} = nominal modifier; \texttt{case} = case marking; \texttt{amod} = adjectival modifier}
	\label{fig:Q1_dependency}
\end{figure}

A \textit{part of speech} (PoS) is a category of words with similar grammatical properties. Almost all languages have the PoS tags \textit{noun} and \textit{verb}. PoS tagging is the process of annotating each token in a text with the corresponding PoS tag (see Figure \ref{fig:Q1_dependency}). The tagging is based both on the token itself and its context. Therefore, it is necessary to first tokenize the text and identify end-of-sentence punctuation. More advanced NLP technologies use the information produced by the PoS tagger. For example, both lemmatization and dependency tree parsing of Stanford CoreNLP (\cite{CoreNLP2014}) have the requirement for PoS tagging.

\subsection{Stemming / Lemmatization}
\label{sec:stemming_lemmatization}

The goal of both stemming and lemmatization is to reduce inflectional and derivationally related forms of a word to a common base form.

\textbf{Stemming} reduces related words to the same stem (root form of the word) by removing different endings of the words. To achieve that, most stemming algorithms refer to a crude heuristic process that chops off the ends of words. The most common algorithm for stemming in English is the Porter's algorithm \citep{Porter1980}. It is based on simple rules that are applied to the longest suffix. For example, there is a rule `\texttt{ies $\rightarrow$ i}' which means that the suffix `\texttt{ies}' will be reduced to `\texttt{i}'. This is needed for words like `\textit{ponies}' which are reduced to `\textit{poni}'. In addition, there is a rule `\texttt{y $\rightarrow$ i}' which ensures that 'pony' is also reduced to `\textit{poni}'. In the sample world, stemming can be used to ensure that the words `\textit{directors}', `\textit{director}', `\textit{directing}' and `\textit{directed}' can be used to find the table \texttt{Director}, because they are all reduced to the same stem `\textit{direct}'. The disadvantage of stemming is, that the generated stem not only consists of words with a similar meaning. For example, the adjective `\textit{direct}' would be reduced to the same stem as `\textit{director}', but the meaning differs. An example question could be `\textit{Which movie has a direct interaction scene between Brad Pitt and Jessica Alba?}', where the word `\textit{direct}' has nothing to do with the director of the movie. In general, stemming increases recall but harms precision.

\textbf{Lemmatization} removes inflectional endings and returns the lemma, which is either the base form of the word or the dictionary form. To achieve that, lemmatization algorithms usually use a vocabulary and morphological analysis of the words. For example, `\textit{directors}' and `\textit{director}' would both have the lemma `\textit{director}' but `\textit{directing}' and `\textit{directed}' would lead to the verb `\textit{(to) direct}'. Lemmatization is normally used together with PoS tagging, which leads to the distinction between the verb `\textit{direct}' and the adjective `\textit{direct}'. Another example would be the question `\textit{Who wrote ``Inglourious Basterds''?}' where lemmatization can translate the irregular verb `\textit{wrote}' into `\textit{write}'. In contrast to stemming, lemmatization can be improved by using context.

\subsection{Parsing}
\label{sec:parsing}

\begin{figure}
	\includegraphics[width=\columnwidth]{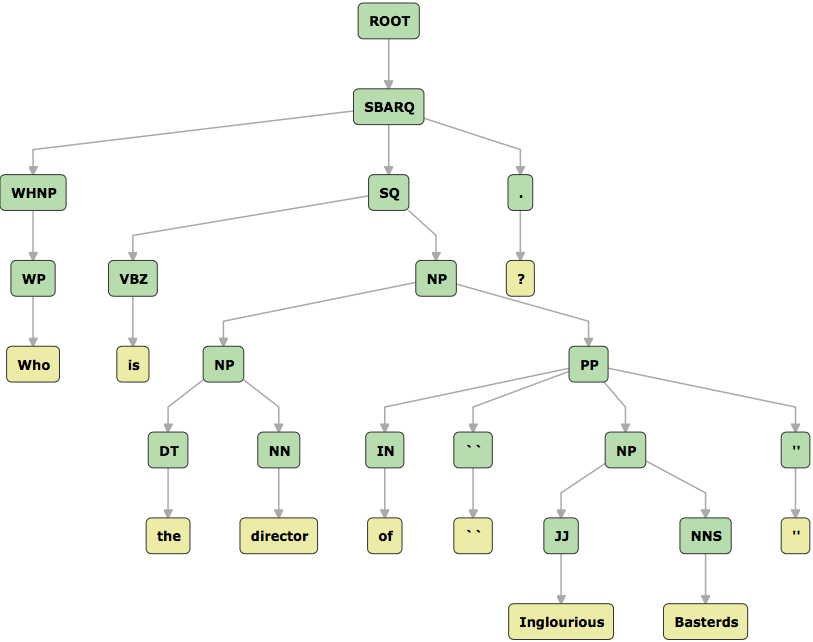}
	\caption{Constituency tree for the sample question \texttt{Q1} using Stanford CoreNLP.}
	\label{fig:Q1_constituency}
\end{figure}

Parsing is the process of analyzing the grammatical structures (syntax) of a sentence. Usually, the parser is based on context-free grammar. There are two main directions of how to look at the syntax: the first main direction, dependency syntax (Figure \ref{fig:Q1_dependency}),  looks at the syntactic structures as relations between words. The other main direction, constituency syntax (Figure \ref{fig:Q1_constituency}), analyses not only the words but also more complex entities (constituents). The information generated through parsing is traditionally stored as a syntax tree. NLIs can use the information on relations found in the syntax tree to generate more accurate queries.

\section{Limitations}
\label{sec:limitations}

This survey's evaluation focuses on the ten sample questions introduced in Section \ref{subsec:input_questions}. Those sample questions are based on the operators of the formal query languages SQL and SPARQL. This leads to the following limitations of the evaluation. 

\textbf{Limitation 1 - Theoretical}: Our evaluation is theoretical and only based on the papers. A few systems have online demos (e.g., SPARKLIS \citep{SPARKLIS2017}), but others --- especially older ones --- are not available any more. Therefore, to handle all systems equally, we based our evaluation fully on the papers of the systems.

\textbf{Limitation 2 - Computational Performance}: We completely ignored the computational performance of the systems. This limitation is based on the different focuses of the systems. For example, \cite{BELA2012} (BELA) propose an approach to increase the speed and computational efficiency. \cite{NLQ/A2017} (NLQ/A) and \cite{QUICK2009} (QUICK) propose algorithms to optimize the number of user interactions for solving ambiguity problems.

\textbf{Limitation 3 - Accuracy}: In our evaluation, we ignored the accuracy given in the papers. There are multiple reasons for this decision (see Table \ref{tab:categorization}): (a) Depending on the focus of the system, not all papers include evaluations based on accuracy. (b) Different usages of the same metrics, for example, \cite{Athena2016} (ATHENA) only consider questions where the system could calculate an answer, but \cite{FREyA2010} (FREyA) evaluate the system against the same dataset but based on needed user interactions. (c) The number of questions in different datasets can be highly different, for example, GeoData250 consists of 250 questions, Task 2 of QALD\footnote{\url{http://qald.aksw.org/}}-4 only of 25. (d) Even if two systems are using the same dataset, two system can preform different pre-processing, for example, \cite{Querix2006} (QUERIX) reduces the GeoData880 to syntactic patterns instead of all questions. All those factors make it impossible to directly compare the systems based on the given evaluation metrics in the paper.

\section{Recently Developed NLIs}
\label{sec:systems}
 
In this section, we will focus on more recent NLIs starting from 2005. We will not discuss about older systems like BASEBALL \citep{BASEBALL1961}, LUNAR \citep{LUNAR1973}, RENDEZVOUS \citep{RENDEZVOUS1974}, LADDER \citep{LADDER1977}, Chat-80 \citep{Chat80}, ASK \citep{ASK1983} and JANUS \citep{JANUS1989}, which are often quoted in the research field of NLIs. We will systematically analyze 24 recently developed systems in Sections \ref{sec:keyword-based} to \ref{sec:grammar-based} based on the sample world introduced in Section \ref{sec:sample_world}. The main goal is to highlight strengths and weaknesses of the different approaches based on a particular data model and particular questions to be able to directly compare the systems. The explanation is based on the information describing the systems found in the papers. Finally, in Section \ref{sec:NLI-summary} we will evaluate the systems against the the sample questions and give an overall interpretation of the system.

There are different ways to classify NLIs. In this survey, we divide the NLIs into four main groups based on the technical approach they use:

\begin{enumerate}
  \item \textbf{Keyword-based systems}\\
	The core of these systems is the lookup step, where the systems try to match the given keywords against an inverted index of the base and meta data. These systems cannot answer aggregation queries like question \texttt{Q7} `\textit{What was the best movie of each genre?}'. The main advantage of this approach is the simplicity and adaptability. 

  \item \textbf{Pattern-based systems}\\
	These systems extend the keyword-based systems with NLP technologies to handle more than keywords and also add natural language patterns. The patterns can be domain-independent or domain-\linebreak dependent. An example for a domain-independent pattern would be allowing aggregations with the words `\textit{by}' or `\textit{how many}'. A domain-dependent pattern could for example be a concept like `\textit{great movie}'.

  \item \textbf{Parsing-based systems}\\
	These systems parse the input question and use the generated information about the structure of the question to understand the grammatical structure. The parse tree contains a lot of information about single tokens, but also about how tokens can be grouped together to form phrases. The main advantage of this approach is that the semantic meaning can be mapped to certain production rules (query generation).

  \item \textbf{Grammar-based systems}\\
	The core of these systems is a set of rules (grammar) that define the questions a user can ask the system. The main advantage of this approach is that the system can give users natural language suggestions during typing their questions. Each question that is formalized this way, can be answered by the system.
\end{enumerate}

Table \ref{tab:categorization} gives an overview of the most representative NLIs of these four categories that we will discuss in Sections \ref{sec:keyword-based} to \ref{sec:grammar-based}. The table also shows which kind of query languages the systems support (e.g., SQL) as well as which NLP technologies are used. Moreover, for each system we report on which data sets they were evaluated. If available, we also show precision and recall. However, we notice that a direct comparison between these systems is very difficult since either they are evaluated on different data sets, they do not report precision/recall measures or they provide user studies without precision/recall information. In summary, our approach of systematically evaluating these systems on a sample world with queries of increasing complexity enables a better comparison of the different approaches. 

In the next subsections, we will systematically analyze the systems in more detail.

\begin{table*}
\centering
\caption{Categorization of the NLIs and the used NLP technologies. (\ding{51}: using; \ding{115}: partly using; \ding{55}: not using; ?: not documented). Data set abbreviations: EBM (Evidence Based Medicine), FinDWH (Financial Data Warehouse), IMDB (Internet MovieDB), MAS (Mircosoft Academic Search), QALD (Question Answering Over Linked Data).}
\label{tab:categorization}
\begin{tabular}{lll|ccccccc|ccc|lll}
 & & & 
 \begin{turn}{90}\textbf{Stop word}\end{turn} & 
 \begin{turn}{90}\textbf{Synonym}\end{turn} & 
 \begin{turn}{90}\textbf{PoS}\end{turn} & 
 \begin{turn}{90}\textbf{Stem / Lemma}\end{turn} & 
 \begin{turn}{90}\textbf{NER}\end{turn} & 
 \begin{turn}{90}\textbf{Dep. Parser}\end{turn} & 
 \begin{turn}{90}\textbf{Const. Parser}\end{turn} &
 \begin{turn}{90}\textbf{SQL}\end{turn} & 
 \begin{turn}{90}\textbf{SPARQL}\end{turn} & 
 \begin{turn}{90}\textbf{Other}\end{turn}  & 
 \begin{turn}{90}\textbf{Precision}\end{turn}  & 
 \begin{turn}{90}\textbf{Recall}\end{turn}  & 
 \textbf{Dataset} \\ 
 \hline
\multirow{7}{*}{Keyword} & \textbf{SODA} & \textit{2012} & \cellcolor[HTML]{CB0000}{\color[HTML]{FFFFFF} \ding{55}} & \cellcolor[HTML]{009901}{\color[HTML]{FFFFFF} \ding{51}} & \cellcolor[HTML]{CB0000}{\color[HTML]{FFFFFF} \ding{55}} & \cellcolor[HTML]{CB0000}{\color[HTML]{FFFFFF} \ding{55}} & \cellcolor[HTML]{CB0000}{\color[HTML]{FFFFFF} \ding{55}} & \cellcolor[HTML]{CB0000}{\color[HTML]{FFFFFF} \ding{55}} & \cellcolor[HTML]{CB0000}{\color[HTML]{FFFFFF} \ding{55}} & \cellcolor[HTML]{009901}{\color[HTML]{FFFFFF} \ding{51}} & \cellcolor[HTML]{CB0000}{\color[HTML]{FFFFFF} \ding{55}} & \cellcolor[HTML]{CB0000}{\color[HTML]{FFFFFF} \ding{55}} & 0.82 & 0.76 & 13 queries on FinDWH \\
& \textbf{NLP-Reduce} & \textit{2007} & 
\cellcolor[HTML]{CB0000}{\color[HTML]{FFFFFF} \ding{55}}  & \cellcolor[HTML]{009901}{\color[HTML]{FFFFFF} \ding{51}} & \cellcolor[HTML]{CB0000}{\color[HTML]{FFFFFF} \ding{55}} & \cellcolor[HTML]{009901}{\color[HTML]{FFFFFF} \ding{51}} & \cellcolor[HTML]{CB0000}{\color[HTML]{FFFFFF} \ding{55}} & \cellcolor[HTML]{CB0000}{\color[HTML]{FFFFFF} \ding{55}} & \cellcolor[HTML]{CB0000}{\color[HTML]{FFFFFF} \ding{55}}& \cellcolor[HTML]{009901}{\color[HTML]{FFFFFF} \ding{51}} & \cellcolor[HTML]{CB0000}{\color[HTML]{FFFFFF} \ding{55}} & \cellcolor[HTML]{009901}{\color[HTML]{FFFFFF} \ding{51}} & 0.71 & 0.76 & GeoData880\\
& \textbf{Pr\'{e}cis} & \textit{2008} & 
\cellcolor[HTML]{CB0000}{\color[HTML]{FFFFFF} \ding{55}} & \cellcolor[HTML]{CB0000}{\color[HTML]{FFFFFF} \ding{55}} & \cellcolor[HTML]{CB0000}{\color[HTML]{FFFFFF} \ding{55}} & \cellcolor[HTML]{CB0000}{\color[HTML]{FFFFFF} \ding{55}} & \cellcolor[HTML]{CB0000}{\color[HTML]{FFFFFF} \ding{55}} & \cellcolor[HTML]{CB0000}{\color[HTML]{FFFFFF} \ding{55}} & \cellcolor[HTML]{CB0000}{\color[HTML]{FFFFFF} \ding{55}} & 
\cellcolor[HTML]{009901}{\color[HTML]{FFFFFF} \ding{51}} & \cellcolor[HTML]{CB0000}{\color[HTML]{FFFFFF} \ding{55}} & \cellcolor[HTML]{CB0000}{\color[HTML]{FFFFFF} \ding{55}} & - & - & user study on IMDb \\
& \textbf{QUICK} & \textit{2009} & 
\cellcolor[HTML]{CB0000}{\color[HTML]{FFFFFF} \ding{55}} & \cellcolor[HTML]{009901}{\color[HTML]{FFFFFF} \ding{51}} & \cellcolor[HTML]{CB0000}{\color[HTML]{FFFFFF} \ding{55}} & \cellcolor[HTML]{CB0000}{\color[HTML]{FFFFFF} \ding{55}} & \cellcolor[HTML]{CB0000}{\color[HTML]{FFFFFF} \ding{55}} & \cellcolor[HTML]{CB0000}{\color[HTML]{FFFFFF} \ding{55}} & \cellcolor[HTML]{CB0000}{\color[HTML]{FFFFFF} \ding{55}} & 
\cellcolor[HTML]{CB0000}{\color[HTML]{FFFFFF} \ding{55}} & \cellcolor[HTML]{009901}{\color[HTML]{FFFFFF} \ding{51}} & \cellcolor[HTML]{CB0000}{\color[HTML]{FFFFFF} \ding{55}} & - & - & 100 queries on IMDb \\

& \textbf{QUEST} & \textit{2013} & 
\cellcolor[HTML]{C0C0C0}{\color[HTML]{FFFFFF} \textbf{?}} & \cellcolor[HTML]{C0C0C0}{\color[HTML]{FFFFFF} \textbf{?}} & \cellcolor[HTML]{C0C0C0}{\color[HTML]{FFFFFF} \textbf{?}} & \cellcolor[HTML]{C0C0C0}{\color[HTML]{FFFFFF} \textbf{?}} & \cellcolor[HTML]{CB0000}{\color[HTML]{FFFFFF} \ding{55}} & \cellcolor[HTML]{CB0000}{\color[HTML]{FFFFFF} \ding{55}} & \cellcolor[HTML]{CB0000}{\color[HTML]{FFFFFF} \ding{55}} & 
\cellcolor[HTML]{009901}{\color[HTML]{FFFFFF} \ding{51}} & \cellcolor[HTML]{CB0000}{\color[HTML]{FFFFFF} \ding{55}} & \cellcolor[HTML]{CB0000}{\color[HTML]{FFFFFF} \ding{55}} & - & - & no eval, demo on IMDb\\
& \textbf{SINA} & \textit{2015} & 
\cellcolor[HTML]{009901}{\color[HTML]{FFFFFF} \ding{51}} & \cellcolor[HTML]{CB0000}{\color[HTML]{FFFFFF} \ding{55}} & \cellcolor[HTML]{CB0000}{\color[HTML]{FFFFFF} \ding{55}} & \cellcolor[HTML]{009901}{\color[HTML]{FFFFFF} \ding{51}} & \cellcolor[HTML]{CB0000}{\color[HTML]{FFFFFF} \ding{55}} & \cellcolor[HTML]{CB0000}{\color[HTML]{FFFFFF} \ding{55}} & \cellcolor[HTML]{CB0000}{\color[HTML]{FFFFFF} \ding{55}} &  
\cellcolor[HTML]{CB0000}{\color[HTML]{FFFFFF} \ding{55}} & \cellcolor[HTML]{009901}{\color[HTML]{FFFFFF} \ding{51}} & \cellcolor[HTML]{CB0000}{\color[HTML]{FFFFFF} \ding{55}} & 0.32 & 0.32 & QALD-3\\
& \textbf{Aqqu} & \textit{2015} & 
\cellcolor[HTML]{C0C0C0}{\color[HTML]{FFFFFF} \textbf{?}} & \cellcolor[HTML]{009901}{\color[HTML]{FFFFFF} \ding{51}} & \cellcolor[HTML]{009901}{\color[HTML]{FFFFFF} \ding{51}} & \cellcolor[HTML]{009901}{\color[HTML]{FFFFFF} \ding{51}} & \cellcolor[HTML]{009901}{\color[HTML]{FFFFFF} \ding{51}} & \cellcolor[HTML]{CB0000}{\color[HTML]{FFFFFF} \ding{55}} & \cellcolor[HTML]{CB0000}{\color[HTML]{FFFFFF} \ding{55}} & 
\cellcolor[HTML]{CB0000}{\color[HTML]{FFFFFF} \ding{55}} & \cellcolor[HTML]{009901}{\color[HTML]{FFFFFF} \ding{51}} & \cellcolor[HTML]{CB0000}{\color[HTML]{FFFFFF} \ding{55}} & - & - & \makecell[l]{Free917 (76.4\% accuracy) \\ WebQuestions (49.4\% F1)}\\ 
\hline
\multirow{2}{*}{Pattern} & \textbf{NLQ/A} & \textit{2017} & \cellcolor[HTML]{009901}{\color[HTML]{FFFFFF} \ding{51}} & \cellcolor[HTML]{009901}{\color[HTML]{FFFFFF} \ding{51}} & \cellcolor[HTML]{CB0000}{\color[HTML]{FFFFFF} \ding{55}} & \cellcolor[HTML]{009901}{\color[HTML]{FFFFFF} \ding{51}} & \cellcolor[HTML]{CB0000}{\color[HTML]{FFFFFF} \ding{55}} & \cellcolor[HTML]{CB0000}{\color[HTML]{FFFFFF} \ding{55}} & \cellcolor[HTML]{CB0000}{\color[HTML]{FFFFFF} \ding{55}} & \cellcolor[HTML]{CB0000}{\color[HTML]{FFFFFF} \ding{55}} & \cellcolor[HTML]{009901}{\color[HTML]{FFFFFF} \ding{51}} & \cellcolor[HTML]{CB0000}{\color[HTML]{FFFFFF} \ding{55}} & 0.84 & 0.78 & QALD-5 \\
& \textbf{QuestIO} & \textit{2008} & 
\cellcolor[HTML]{CB0000}{\color[HTML]{FFFFFF} \ding{55}} & \cellcolor[HTML]{C0C0C0}{\color[HTML]{FFFFFF} \textbf{?}} & \cellcolor[HTML]{009901}{\color[HTML]{FFFFFF} \ding{51}} & \cellcolor[HTML]{009901}{\color[HTML]{FFFFFF} \ding{51}} & \cellcolor[HTML]{CB0000}{\color[HTML]{FFFFFF} \ding{55}} & \cellcolor[HTML]{CB0000}{\color[HTML]{FFFFFF} \ding{55}} & \cellcolor[HTML]{CB0000}{\color[HTML]{FFFFFF} \ding{55}} & 
\cellcolor[HTML]{CB0000}{\color[HTML]{FFFFFF} \ding{55}} & \cellcolor[HTML]{009901}{\color[HTML]{FFFFFF} \ding{51}} & \cellcolor[HTML]{CB0000}{\color[HTML]{FFFFFF} \ding{55}} & 0.72 & 0.72 & 22 GATE questions\\ 
\hline
\multirow{8}{*}{Parsing} & \textbf{ATHENA} & \textit{2016} & \cellcolor[HTML]{CB0000}{\color[HTML]{FFFFFF} \ding{55}} & \cellcolor[HTML]{009901}{\color[HTML]{FFFFFF} \ding{51}} & \cellcolor[HTML]{009901}{\color[HTML]{FFFFFF} \ding{51}} & \cellcolor[HTML]{009901}{\color[HTML]{FFFFFF} \ding{51}} & \cellcolor[HTML]{009901}{\color[HTML]{FFFFFF} \ding{51}} & \cellcolor[HTML]{009901}{\color[HTML]{FFFFFF} \ding{51}} & \cellcolor[HTML]{CB0000}{\color[HTML]{FFFFFF} \ding{55}} &  \cellcolor[HTML]{009901}{\color[HTML]{FFFFFF} \ding{51}} & \cellcolor[HTML]{CB0000}{\color[HTML]{FFFFFF} \ding{55}} & \cellcolor[HTML]{CB0000}{\color[HTML]{FFFFFF} \ding{55}} & 1.0 & 0.87 & GeoData250\\
& \textbf{Querix} & \textit{2006} & 
\cellcolor[HTML]{CB0000}{\color[HTML]{FFFFFF} \ding{55}} & \cellcolor[HTML]{009901}{\color[HTML]{FFFFFF} \ding{51}} & \cellcolor[HTML]{009901}{\color[HTML]{FFFFFF} \ding{51}} & \cellcolor[HTML]{C0C0C0}{\color[HTML]{FFFFFF} \textbf{?}} & \cellcolor[HTML]{CB0000}{\color[HTML]{FFFFFF} \ding{55}} & \cellcolor[HTML]{CB0000}{\color[HTML]{FFFFFF} \ding{55}} & \cellcolor[HTML]{009901}{\color[HTML]{FFFFFF} \ding{51}} & 
\cellcolor[HTML]{CB0000}{\color[HTML]{FFFFFF} \ding{55}} & \cellcolor[HTML]{009901}{\color[HTML]{FFFFFF} \ding{51}} & \cellcolor[HTML]{CB0000}{\color[HTML]{FFFFFF} \ding{55}} & 0.78 & 0.79 & GeoData880\\
& \textbf{FREyA} & \textit{2010} & 
\cellcolor[HTML]{C0C0C0}{\color[HTML]{FFFFFF} \textbf{?}} & \cellcolor[HTML]{009901}{\color[HTML]{FFFFFF} \ding{51}} & \cellcolor[HTML]{009901}{\color[HTML]{FFFFFF} \ding{51}} & \cellcolor[HTML]{C0C0C0}{\color[HTML]{FFFFFF} \textbf{?}} & \cellcolor[HTML]{CB0000}{\color[HTML]{FFFFFF} \ding{55}} & \cellcolor[HTML]{CB0000}{\color[HTML]{FFFFFF} \ding{55}} & \cellcolor[HTML]{009901}{\color[HTML]{FFFFFF} \ding{51}} & 
\cellcolor[HTML]{CB0000}{\color[HTML]{FFFFFF} \ding{55}} & \cellcolor[HTML]{009901}{\color[HTML]{FFFFFF} \ding{51}} & \cellcolor[HTML]{CB0000}{\color[HTML]{FFFFFF} \ding{55}} & 0.92 & 0.92 & GeoData250\\ 
& \textbf{BELA} & \textit{2012} & 
\cellcolor[HTML]{CB0000}{\color[HTML]{FFFFFF} \ding{55}} & \cellcolor[HTML]{009901}{\color[HTML]{FFFFFF} \ding{51}} & \cellcolor[HTML]{009901}{\color[HTML]{FFFFFF} \ding{51}} & \cellcolor[HTML]{C0C0C0}{\color[HTML]{FFFFFF} \textbf{?}} & \cellcolor[HTML]{CB0000}{\color[HTML]{FFFFFF} \ding{55}} & \cellcolor[HTML]{CB0000}{\color[HTML]{FFFFFF} \ding{55}} & \cellcolor[HTML]{009901}{\color[HTML]{FFFFFF} \ding{51}} & 
\cellcolor[HTML]{CB0000}{\color[HTML]{FFFFFF} \ding{55}} & \cellcolor[HTML]{009901}{\color[HTML]{FFFFFF} \ding{51}} & \cellcolor[HTML]{CB0000}{\color[HTML]{FFFFFF} \ding{55}} & 0.62 & 0.73 & QALD-2\\
& \textbf{USI Answers} & \textit{2013} & 
\cellcolor[HTML]{C0C0C0}{\color[HTML]{FFFFFF} \textbf{?}} & \cellcolor[HTML]{009901}{\color[HTML]{FFFFFF} \ding{51}} & \cellcolor[HTML]{009901}{\color[HTML]{FFFFFF} \ding{51}} & \cellcolor[HTML]{009901}{\color[HTML]{FFFFFF} \ding{51}} & \cellcolor[HTML]{009901}{\color[HTML]{FFFFFF} \ding{51}} & \cellcolor[HTML]{009901}{\color[HTML]{FFFFFF} \ding{51}} & \cellcolor[HTML]{CB0000}{\color[HTML]{FFFFFF} \ding{55}} & 
\cellcolor[HTML]{009901}{\color[HTML]{FFFFFF} \ding{51}} & \cellcolor[HTML]{009901}{\color[HTML]{FFFFFF} \ding{51}} & \cellcolor[HTML]{009901}{\color[HTML]{FFFFFF} \ding{51}} & 0.76 & 0.94 & Enterprise data\\
& \textbf{NaLIX} & \textit{2005} & 
\cellcolor[HTML]{CB0000}{\color[HTML]{FFFFFF} \ding{55}} & \cellcolor[HTML]{009901}{\color[HTML]{FFFFFF} \ding{51}} & \cellcolor[HTML]{009901}{\color[HTML]{FFFFFF} \ding{51}} & \cellcolor[HTML]{C0C0C0}{\color[HTML]{FFFFFF} \textbf{?}} & \cellcolor[HTML]{CB0000}{\color[HTML]{FFFFFF} \ding{55}} & \cellcolor[HTML]{CB0000}{\color[HTML]{FFFFFF} \ding{55}} & \cellcolor[HTML]{009901}{\color[HTML]{FFFFFF} \ding{51}} & 
\cellcolor[HTML]{CB0000}{\color[HTML]{FFFFFF} \ding{55}} & \cellcolor[HTML]{CB0000}{\color[HTML]{FFFFFF} \ding{55}} & \cellcolor[HTML]{009901}{\color[HTML]{FFFFFF} \ding{51}} & - & - & no eval, demo paper\\
& \textbf{NaLIR} & \textit{2014} & 
\cellcolor[HTML]{CB0000}{\color[HTML]{FFFFFF} \ding{55}} & \cellcolor[HTML]{009901}{\color[HTML]{FFFFFF} \ding{51}} & \cellcolor[HTML]{009901}{\color[HTML]{FFFFFF} \ding{51}} & \cellcolor[HTML]{C0C0C0}{\color[HTML]{FFFFFF} \textbf{?}} & \cellcolor[HTML]{CB0000}{\color[HTML]{FFFFFF} \ding{55}} & \cellcolor[HTML]{CB0000}{\color[HTML]{FFFFFF} \ding{55}} & \cellcolor[HTML]{009901}{\color[HTML]{FFFFFF} \ding{51}} & 
\cellcolor[HTML]{009901}{\color[HTML]{FFFFFF} \ding{51}} & \cellcolor[HTML]{CB0000}{\color[HTML]{FFFFFF} \ding{55}} & \cellcolor[HTML]{CB0000}{\color[HTML]{FFFFFF} \ding{55}} & - & - & user study on MAS\\
& \textbf{BioSmart} & \textit{2017} & 
\cellcolor[HTML]{C0C0C0}{\color[HTML]{FFFFFF} \textbf{?}} & \cellcolor[HTML]{C0C0C0}{\color[HTML]{FFFFFF} \textbf{?}} & \cellcolor[HTML]{009901}{\color[HTML]{FFFFFF} \ding{51}} & \cellcolor[HTML]{C0C0C0}{\color[HTML]{FFFFFF} \textbf{?}} & \cellcolor[HTML]{CB0000}{\color[HTML]{FFFFFF} \ding{55}} & \cellcolor[HTML]{CB0000}{\color[HTML]{FFFFFF} \ding{55}} & \cellcolor[HTML]{009901}{\color[HTML]{FFFFFF} \ding{51}} & 
\cellcolor[HTML]{009901}{\color[HTML]{FFFFFF} \ding{51}} & \cellcolor[HTML]{CB0000}{\color[HTML]{FFFFFF} \ding{55}} & \cellcolor[HTML]{009901}{\color[HTML]{FFFFFF} \ding{51}} & - & - & no eval, biodata\\ 
\hline
\multirow{7}{*}{Grammar} & \textbf{TR Discover} & \textit{2015} & \cellcolor[HTML]{C0C0C0}{\color[HTML]{FFFFFF} \textbf{?}} & \cellcolor[HTML]{F8A102}{\color[HTML]{FFFFFF} \ding{115}} & \cellcolor[HTML]{CB0000}{\color[HTML]{FFFFFF} \ding{55}} & \cellcolor[HTML]{CB0000}{\color[HTML]{FFFFFF} \ding{55}} & \cellcolor[HTML]{CB0000}{\color[HTML]{FFFFFF} \ding{55}} & \cellcolor[HTML]{CB0000}{\color[HTML]{FFFFFF} \ding{55}} & \cellcolor[HTML]{CB0000}{\color[HTML]{FFFFFF} \ding{55}} & \cellcolor[HTML]{009901}{\color[HTML]{FFFFFF} \ding{51}} & \cellcolor[HTML]{009901}{\color[HTML]{FFFFFF} \ding{51}} & \cellcolor[HTML]{CB0000}{\color[HTML]{FFFFFF} \ding{55}} & 0.75 & 0.84 & QALD-4 (Task 2)\\
& \textbf{Ginseng} & \textit{2005} & 
\cellcolor[HTML]{CB0000}{\color[HTML]{FFFFFF} \ding{55}} & \cellcolor[HTML]{009901}{\color[HTML]{FFFFFF} \ding{51}} & \cellcolor[HTML]{CB0000}{\color[HTML]{FFFFFF} \ding{55}} & \cellcolor[HTML]{CB0000}{\color[HTML]{FFFFFF} \ding{55}} & \cellcolor[HTML]{CB0000}{\color[HTML]{FFFFFF} \ding{55}} & \cellcolor[HTML]{CB0000}{\color[HTML]{FFFFFF} \ding{55}} & \cellcolor[HTML]{CB0000}{\color[HTML]{FFFFFF} \ding{55}} & 
\cellcolor[HTML]{CB0000}{\color[HTML]{FFFFFF} \ding{55}} & \cellcolor[HTML]{009901}{\color[HTML]{FFFFFF} \ding{51}} & \cellcolor[HTML]{CB0000}{\color[HTML]{FFFFFF} \ding{55}} & 0.93 & 0.98 & GeoData880 \\
& \textbf{SQUALL} & \textit{2014} & 
\cellcolor[HTML]{CB0000}{\color[HTML]{FFFFFF} \ding{55}} & \cellcolor[HTML]{009901}{\color[HTML]{FFFFFF} \ding{51}} & \cellcolor[HTML]{009901}{\color[HTML]{FFFFFF} \ding{51}} & \cellcolor[HTML]{CB0000}{\color[HTML]{FFFFFF} \ding{55}} & \cellcolor[HTML]{CB0000}{\color[HTML]{FFFFFF} \ding{55}} & \cellcolor[HTML]{CB0000}{\color[HTML]{FFFFFF} \ding{55}} & \cellcolor[HTML]{009901}{\color[HTML]{FFFFFF} \ding{51}} & 
\cellcolor[HTML]{CB0000}{\color[HTML]{FFFFFF} \ding{55}} & \cellcolor[HTML]{009901}{\color[HTML]{FFFFFF} \ding{51}} & \cellcolor[HTML]{CB0000}{\color[HTML]{FFFFFF} \ding{55}} & 0.93 & 0.88 & QALD-3\\
& \textbf{MEANS} & \textit{2015} & 
\cellcolor[HTML]{009901}{\color[HTML]{FFFFFF} \ding{51}} & \cellcolor[HTML]{009901}{\color[HTML]{FFFFFF} \ding{51}} & \cellcolor[HTML]{009901}{\color[HTML]{FFFFFF} \ding{51}} & \cellcolor[HTML]{009901}{\color[HTML]{FFFFFF} \ding{51}} & \cellcolor[HTML]{009901}{\color[HTML]{FFFFFF} \ding{51}} & \cellcolor[HTML]{CB0000}{\color[HTML]{FFFFFF} \ding{55}} & \cellcolor[HTML]{CB0000}{\color[HTML]{FFFFFF} \ding{55}} & 
\cellcolor[HTML]{CB0000}{\color[HTML]{FFFFFF} \ding{55}} & \cellcolor[HTML]{009901}{\color[HTML]{FFFFFF} \ding{51}} & \cellcolor[HTML]{CB0000}{\color[HTML]{FFFFFF} \ding{55}} & 0.65 & - & \makecell[l]{EBM}\\
& \textbf{AskNow} & \textit{2016} & 
\cellcolor[HTML]{CB0000}{\color[HTML]{FFFFFF} \ding{55}} & \cellcolor[HTML]{009901}{\color[HTML]{FFFFFF} \ding{51}} & \cellcolor[HTML]{009901}{\color[HTML]{FFFFFF} \ding{51}} & \cellcolor[HTML]{009901}{\color[HTML]{FFFFFF} \ding{51}} & \cellcolor[HTML]{009901}{\color[HTML]{FFFFFF} \ding{51}} & \cellcolor[HTML]{CB0000}{\color[HTML]{FFFFFF} \ding{55}} & \cellcolor[HTML]{CB0000}{\color[HTML]{FFFFFF} \ding{55}} & 
\cellcolor[HTML]{CB0000}{\color[HTML]{FFFFFF} \ding{55}} & \cellcolor[HTML]{009901}{\color[HTML]{FFFFFF} \ding{51}} & \cellcolor[HTML]{CB0000}{\color[HTML]{FFFFFF} \ding{55}} & 0.60 & 0.63 & QALD-5\\
& \textbf{SPARKLIS} & \textit{2017}  & 
\cellcolor[HTML]{CB0000}{\color[HTML]{FFFFFF} \ding{55}} & \cellcolor[HTML]{CB0000}{\color[HTML]{FFFFFF} \ding{55}} & \cellcolor[HTML]{CB0000}{\color[HTML]{FFFFFF} \ding{55}} & \cellcolor[HTML]{CB0000}{\color[HTML]{FFFFFF} \ding{55}} & \cellcolor[HTML]{CB0000}{\color[HTML]{FFFFFF} \ding{55}} & \cellcolor[HTML]{CB0000}{\color[HTML]{FFFFFF} \ding{55}} & \cellcolor[HTML]{CB0000}{\color[HTML]{FFFFFF} \ding{55}} & 
\cellcolor[HTML]{CB0000}{\color[HTML]{FFFFFF} \ding{55}} & \cellcolor[HTML]{009901}{\color[HTML]{FFFFFF} \ding{51}} & \cellcolor[HTML]{CB0000}{\color[HTML]{FFFFFF} \ding{55}} & - & - & user study on QALD-3 \\
& \textbf{GFMed} & \textit{2017} & \cellcolor[HTML]{C0C0C0}{\color[HTML]{FFFFFF} \textbf{?}} & \cellcolor[HTML]{C0C0C0}{\color[HTML]{FFFFFF} \textbf{?}} & \cellcolor[HTML]{C0C0C0}{\color[HTML]{FFFFFF} \textbf{?}} & \cellcolor[HTML]{C0C0C0}{\color[HTML]{FFFFFF} \textbf{?}} & \cellcolor[HTML]{C0C0C0}{\color[HTML]{FFFFFF} \textbf{?}} & \cellcolor[HTML]{C0C0C0}{\color[HTML]{FFFFFF} \textbf{?}} & \cellcolor[HTML]{C0C0C0}{\color[HTML]{FFFFFF} \textbf{?}}& 
\cellcolor[HTML]{CB0000}{\color[HTML]{FFFFFF} \ding{55}} & \cellcolor[HTML]{009901}{\color[HTML]{FFFFFF} \ding{51}} & \cellcolor[HTML]{CB0000}{\color[HTML]{FFFFFF} \ding{55}} & 1 & 0.99 & QALD-4 (Task 2)
\end{tabular}
\end{table*}

\subsection{Keyword-based systems}
\label{sec:keyword-based}
The core of keyword-based NLIs is their lookup step. In this step, the system tries to match the given keywords against an inverted index of the base and meta data. To identify keywords in the input question, some systems are using stop word removal (e.g., NLP-Reduce \citep{NLPReduce2007}), others are expecting only keywords from the users as input (e.g., SODA \citep{SODA2012}). 

Most questions are easily formulated with keywords. However, there are some cases where keywords are not enough to express the intention of the users. For example, for the question `\textit{What was the best movie of each genre?}' (\texttt{Q7}) the ``keyword-only version'' would be something like `\textit{best movie genre}', which is more likely to be interpreted as `\textit{the genre of the best movie}'. If the users would write the question like `\textit{best movie by genre}', a keyword-based NLI would try to lookup the token `\textit{by}' in the base and meta data or classify `\textit{by}' as a stop word and ignore it.

In the following, we will summarize seven keyword-based NLI. We decided to describe SODA \citep{SODA2012} - as the first system - in depth, because it can solve the most of our sample input questions in this category (see Section \ref{subsec:input_questions}). SODA is an NLI that expects only keywords from the user and can handle aggregations by using specific non-natural language templates. Afterwards, the other systems are summarized and we highlight the difference between them to SODA and each other.

\subsubsection{SODA (Search Over DAta warehouse)}
SODA \citep{SODA2012} is a system that provides a keyword-based NLI for relational databases with some extensions in the direction of a pattern-based system. The base data consists of the relational database. The meta data can include multiple ontologies, which are handled like natural language patterns. For example, domain specific ontologies with concepts (like the concept `\textit{great movie}' in the sample world) or DBpedia to identify homonyms and synonyms. SODA uses both inverted indexes (base and meta data) as the basis for finding query matches in the data. The key innovation of SODA is that it provides the possibility to define meta data patterns which specify conceptual models. The concept `\textit{good movie}' could depend on various variables not only on the rating, but for example, also on the number of ratings. The users can then apply this concept to their input questions, for example, they could search for `\textit{all great movie}' (\texttt{Q6}) without having to specify what a great movie is.

Assuming the users want to know the director of the movie `\textit{Inglourious Basterds}' (\texttt{Q1}), the input question for SODA could be: `\textit{director Inglourious Basterds}'.

\begin{figure}
	\includegraphics[width=\columnwidth]{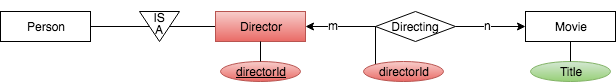}
	\caption{Nodes in the meta data graph corresponding to the keywords `\textit{director}' (red) and `\textit{Inglourious Basterds}' (green) found during the lookup step of SODA.}
	\label{fig:SODA_lookup}
\end{figure}

SODA uses five steps to translate this keyword-based input question into a SQL query. The first step is the lookup: it checks the keywords against the inverted indexes over the database and provides all the nodes in the meta data graph where these keywords are found. For the input question \texttt{Q1}, this means that the keyword `\textit{director}' can be found in the inverted index of the meta data, either the table name \texttt{Director} or to the attribute name \texttt{Director.directorId} and \texttt{Directing.director}-\texttt{Id} (Figure \ref{fig:SODA_lookup}: red). The keyword `\textit{Inglourious Basterds}' is only found in the inverted index of the base data as a value of the attribute \texttt{Movie.Title} (Figure \ref{fig:SODA_lookup}: green). This leads to three different solution sets for the next steps: \{\texttt{Directing.directorId}, \texttt{Movie.Title}\}, \linebreak
\{\texttt{Director.directorId}, \texttt{Movie.Title}\} and \linebreak
\{\texttt{Director}, \texttt{Movie.Title}\}.

The second step is to assign a score to each solution of the lookup step. SODA uses a simple heuristic method, for example in-domain solutions receive a higher score. For the input question, the solution \{\texttt{Director}, \texttt{Movie.Title}\} receives the highest score, because the table name \texttt{Director} it is a full match and not only a fuzzy match like in \texttt{directorId}. Afterwards, only the best n solutions are provided to the next step. 

The third step identifies which tables are used for each of the solutions provided by the previous step. Also, the relationships and inheritance structures between those tables are discovered  in this step. For the best solution of the input question, the tables \texttt{Director} and \texttt{Movie} correspond to the different entry points. An entry point is a node in the meta data graph. The table \texttt{Director} is a child of the table \texttt{Person} (ISA-relationship). Therefore, SODA includes the table \texttt{Per}-\texttt{son} in the solution. To link the table \texttt{Movie} to the other two tables, it is necessary to add more tables to the solution. The closest link is through the table \texttt{Directing} (see Figure \ref{fig:SODA_lookup}) and therefore this table is included.

The fourth step collects the filters. There are two types of filters which are collected. The first one are filters in the input question like `\textit{Inglourious Basterds}'. The second one are filter conditions that occur during traversing the meta data graph like the concept `\textit{great movie}'. 

The fifth and last step generates a \textit{reasonable} and \textit{executable} SQL query from the information collected in the previous steps. A \textit{reasonable} SQL query is a query which considers foreign keys and inheritance patterns in the schema. An \textit{executable} SQL query is a query that can be executed on the underlying database.

The strengths of SODA are the use of meta data patterns and domain ontologies, which allow one to define concepts and include domain specific knowledge. In addition, the inclusion of external sources like DBpedia for homonyms and synonyms is beneficial for finding meaningful results. Furthermore, SODA is designed to evolve and thus improve over time based on user feedback. 

The weaknesses of SODA are that it uses simple word recognition for comparison operators. For example, to retrieve all movies with a rating greater than~9, the input question needs to be written like `\textit{rating~$>$~9}' (\texttt{Q2}). Moreover, SODA uses a very strict syntax for aggregation operators. For example, to retrieve the number of movies per year, the input question needs to be written like `\textit{select count (movie) group by (year)}'. These patterns are useful, but are not in natural language. Furthermore, there is no lemmatization, stemming or any other preprocessing of the input question which can lead to a problem with words that are used in plural. For example the input question `\textit{all movies}' would not detect the table \texttt{Movie} but the input question 'all movie' would display the expected result.

\cite{SODA2012} suggest extending SODA to handle temporal aspects of the data warehouse (e.g., bi-temporal historization). They also pointed out that the GUI of SODA should be improved so that the users are engaged in selecting and ranking the different results. Furthermore, the user feedback provided by SODA is currently very basic and needs to be improved.

\subsubsection{NLP-Reduce}
NLP-Reduce \citep{NLPReduce2007} uses a few simple NLP technologies to `reduce' the input tokens before the lookup in the \textit{Knowledge Base} (KB) based on RDF. The system takes the input question, reduces it to keywords, translates it into SPARQL to query the KB and then returns the result to the user.

In contrast to SODA, NLP-Reduce uses not only synonyms but also two other NLP technologies: (a) stop words and punctuation marks removal and (b) stemming. With the removal of stop words and punctuation marks, NLP-Reduce is able to  answer some fragment and full sentence questions. NLP-Reduce still cannot answer questions with aggregations like `\textit{What was the best movie of each genre?}' (\texttt{Q7}), because it will remove the token `\textit{of each}' as a stop word. Furthermore, stemming helps the user to formulate questions like `\textit{all movies}' which is more natural than `\textit{all movie}' for SODA.

After reducing the input question, NLP-Reduce has similar steps to SODA: (1) search for triples in the RDF graph (similar to base and meta data), where at least one of the question tokens occurs and rank the found triples, (2) search for properties that can be joined with the triples found in the previous step by the remaining question tokens, (3) search for data type property values that match the remaining question tokens and \linebreak(4) generate corresponding SPARQL query.

Compared to SODA, the strength of NLP-Reduce is the reduction of the input question such that non-keyword input questions can be answered. Besides the overall weakness of keyword-based NLIs, NLP-Reduce is not able to answer comparison questions like \texttt{Q2}.

\subsubsection{Pr\'{e}cis}
Pr\'{e}cis \citep{Precis2008} is a keyword-based NLI for relational databases, which supports multiple terms combined through the operators \texttt{AND}, \texttt{OR} and \texttt{NOT}. For example, input question `\textit{Show me all drama and comedy movies.}' (\texttt{Q5}) would be formulated as `\textit{``drama'' OR ``comedy''}'. The answer is an entire multi-relation data-base, which is a logical subset of the original data\-base.

First, Pr\'{e}cis transforms the input question into \textit{disjunct normal form} (DNF). Afterwards, each term in the DNF is looked up in the inverted index of the base data. This is different to SODA, where the inverted index includes the meta data. If a term cannot be found, the next steps are not executed. The third step creates the schema of the logical database subset, which represents the answer of the input question. This includes the identification of the necessary join paths.

The strength of Pr\'{e}cis is the ability to use brackets, \texttt{AND}, \texttt{OR} and \texttt{NOT} to define the input question. However, the weaknesses are that this again composes a logical query language, although a simpler one. Furthermore, it can only solve boolean questions, and the input question can only consist of terms which are located in the base data and not in the meta data. For example, the input question `\textit{Who is the director of ``Inglourious Basterds''?}' (\texttt{Q1}) cannot directly be solved because `\textit{director}' is the name of a table and therefore part of the meta data. There is a mechanism included that adds more information to the answer (e.g., the actors, directors etc. to a movie), but then the user would have to search for the director in the answer.

\subsubsection{QUICK (QUery Intent Constructor for Keywords)}
QUICK \citep{QUICK2009} is an NLI that adds the expressiveness of semantic queries to the convenience of keyword-based search. To achieve this, the users start with a keyword question and then are guided through the process of incremental refinement steps to select the question's intention. The system provides the user with an interface that shows the semantic queries as graphs as well as textual form. 

In a first step, QUICK takes the keywords of the input question and compares them against the KB. Each possible interpretation corresponds to a semantic query. For example, the input question `\textit{Brad Pitt}' can either mean `\textit{movies where Brad Pitt played in}', `\textit{movies directed by Brad Pitt}' or `\textit{movies written by Brad Pitt}' (see Figure \ref{fig:sample_world}). In the next step, the system provides the users with the information in such a way that they can select the semantic query which will answer their question. To do so, QUICK provides the users with possible interpretations of each keyword to select from. This is done with a graph as well as a textual form of the semantic query. The textual form is a translation of the SQL query into natural language based on templates. Furthermore, the system orders the keywords in such a way that the user interactions are as few as possible. When the users select the desired semantic query, QUICK executes it and displays the results in the user interface.

The strength of QUICK is the user interaction interface with the optimization for minimal user interaction during the semantic query selection. The weakness of QUICK is that it is limited to acyclic conjunctions of triple patterns.

\subsubsection{QUEST (QUEry generator for STructured sources)}
QUEST \citep{QUEST2013} is a keyword-based NLI to translate input questions into SQL. It combines semantic and statistical ML techniques for the translation.

The first step is to determine how the keywords in the input question correspond to elements of the database (lookup). In contrast to SODA, QUEST uses two \textit{Hidden Markov Models} (HMM) to choose the relevant elements (ranking). The first HMM is a set of heuristic rules. The second HMM is trained with user feedback. The next step is to identify the possible join paths to connect all the relevant elements from the previous step. QUEST selects the most informative join paths (similar to SODA's step 4). The most informative join paths are those that contain tuples in the database. In the third step, QUEST decides which combination of keyword mapping and join path most likely represents the semantics the users had in mind when formulating the keyword question.

The strength of QUEST is the combination of user feedback and a set of heuristic rules during the ranking. This allows the system to learn from the users over time. A weakness of QUEST is that it is not able to handle concepts such as `\textit{good movie}'.

\subsubsection{SINA}
SINA \citep{SINA2015} is a keyword-based NLI that transforms natural language input questions into conjunctive SPARQL queries. It uses a \textit{Hidden Markov Model} to determine the most suitable resource for a given input question from different datasets.

In the first step, SINA reduces the input question to keywords (similar to NLP-Reduce), by using tokenization, lemmatization and stop word removal. In the next step, the keywords are grouped into segments, with respect to the available resources. For example, the keywords `\textit{Inglourious}' and `\textit{Basterds}' would be grouped into one segment based on the match for `\textit{Inglorious Basterds}'. In the third step, the relevant resources are retrieved based on string matching between the segments and the RDF label of the resource. In the following step, the best subset of resources for the given input question is determined (ranking). The fifth step, SINA constructs the SPARQL query using the graph-structure of the database. Finally, the results, retrieved by evaluating the generated SPARQL query, are shown to the users.

The biggest weakness of SINA is that it can only translate into conjunctive SPARQL queries, which reduces the number of answerable questions.

\subsubsection{Aqqu}
Aqqu \citep{Aqqu2015} is an NLI which uses templates to identify possible relations between keywords. At the end of the translation process, ML is used to rank the possible solutions.

To translate the input question into SPARQL, first the entities from the KB that match (possibly overlapping) parts of the input question are identified. The possible parts are identified by using PoS tags. For example, single token parts must be a noun (\texttt{NN}) and proper nouns (\texttt{NNP}) are not allowed to be split (e.g.,~`\textit{Brad Pitt}'). In the next step, Aqqu uses three different templates which define the general relationship between the keywords. Afterwards, Aqqu tries to identify the corresponding relationship. This can either be done with the help of the input question (verbs and adjectives), or with the help of ML which for example can identify abstract relationship like `\textit{born} $\rightarrow$ \textit{birth date}'. The last step is the ranking which is solved with ML. The best result is achieved by using a binary random forest classifier.

The strength of Aqqu is the identification of abstract relationships. The weakness is the limitation of a keyword-based NLI.

\subsection{Pattern-based systems}
\label{sec:pattern-based}
Pattern-based NLIs are an extension of keyword-based systems with natural language patterns to answer more complex questions like concepts (\texttt{Q6}) or aggregations (\texttt{Q7}). For example, the question `\textit{What was the best movie of each genre?}' (\texttt{Q7}) cannot be formulated with keywords only. It needs at least some linking phrase between `\textit{best movie}' and `\textit{genre}', which indicates the aggregation. This could be done with the non-keyword token (trigger word) `\textit{by}' for the aggregation, which will indicate that the right side includes the keywords for the \texttt{group by}-clause and the left side the keywords for the \texttt{select}-clause. The difficulty with trigger words is to find every possible synonym allowed by natural language. For example, an aggregation could be implied with the word `\textit{by}' but also `\textit{of each}' (compare \texttt{Q7}).

In the following, we summarize two pattern-based NLIs. We decided to describe NLQ/A \citep{NLQ/A2017} in depth, because it is based on the idea that the errors made by NLP technologies are not worth the gain of information. Instead, the system is highly dependent on the users' input to solve ambiguity problems, and therefore it focuses on the optimization of the user interaction.

\subsubsection{NLQ/A}
NLQ/A \citep{NLQ/A2017} is an NLI to query a knowledge graph. The system is based on a new approach without NLP technologies like parsers or PoS taggers. The idea being that the errors made by these technologies are not worth the gain of information. For example, a parse tree helps for certain questions like subqueries (e.g., \texttt{Q9}), but if the parse tree is wrong, the system will fail to translate even simpler questions. Instead, NLQ/A lets the users resolve all ambiguity problems, also those which could be solved with PoS tagging or parse trees. To avoid needing too many interaction steps, NLQ/A provides an efficient greedy approach for the interaction process.

Assuming the users want to know the director of the movie `\textit{Inglourious Basterds}' (\texttt{Q1}), the input question could be: `\textit{Who is the director of ``Inglourious Basterds''?}'.

NLQ/A use four steps to answer the input question. The first step is to detect the phrases of the input question. In general, the phrases can be categorized into two types: \textit{independent} and \textit{dependent} phrases. \textit{Independent} phrases are identified with a phrase dictionary. The dictionary consists of variables, aggregations, operators, modifiers and quantifier phrases. To detect \textit{dependent} phrases, most stop words are removed (simplified input question). Some types of words like prepositions are still needed and therefore kept. Next \texttt{1:n}-grams are generated. Phrases starting with prepositions are discarded. After stop word removal, the input question \texttt{Q1} would become `\textit{director of Inglourious Basterds}'. If \texttt{n} is set to~2, the extracted phrases would be: \{`\textit{director}', `\textit{director of}', `\textit{Inglourious}', `\textit{Inglourious Basterds}', `\textit{Basterds}'\}. Next, the phrases are extended according to a synonym dictionary. For example if there is a phrase `\textit{starring}' it would be extended with the phrase `\textit{playing}'. Those extended phrases are mapped to the knowledge graph based on the string similarity (edit distance). For one extended phrase there can be multiple candidate mappings.

\begin{figure}
	\includegraphics[width=\columnwidth]{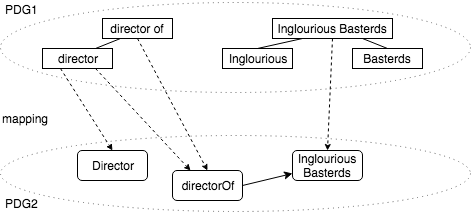}
	\caption{Phrase dependency graph (PDG) for the input question `\textit{Who is the director of ``Inglourious Basterds''?}'. (PDG1: input question; PDG2: knowledge graph)}
	\label{fig:NLQA_PDG}
\end{figure}

The next step takes the candidate mappings and tries to find the true meaning of the input question with the help of the users. To reduce the amount of interactions for the user, a \textit{phrase dependency graph} (PDG) is proposed. The PDG consists of two parts: (PDG1) a graph where each node represents a phrase, two phrases are connected if they share at least one common token and (PDG2) a subgraph of the knowledge graph consisting of the candidates where each node represents a candidate, two nodes are connected if they are adjacent in the knowledge graph. The two parts are connected with edges, representing the mapping between phrases and candidates (see Figure \ref{fig:NLQA_PDG}).

In the third step, the users get involved to solve the ambiguity given in the PDG. In order to reduce the necessary user interactions, the NLI tries to find those edges which resolve the most ambiguities (similar to the idea of QUICK).

\begin{figure}
	\includegraphics[width=\columnwidth]{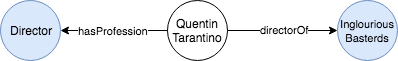}
	\caption{Answer graph generated for the input question `\textit{Who is the director of ``Inglourious Basterds''?}' based on the selected candidates (blue).}
	\label{fig:NLQA_Answer}
\end{figure}

The last step takes the selected candidates and tries to connect them into one graph. The connected graph will include the answer to the question. Groups of already connected candidates in the PDG2 are called query fragments. In Figure \ref{fig:NLQA_PDG}, the candidates `\textit{director-Of}' and `\textit{Inglourious Basterds}' are one query fragment. For each query fragment, the system tries to find the path with the highest similarity to the simplified input question. For the input question \texttt{Q1}, if the users select `\textit{Director}' as candidate in step 3, the system would find the path as shown in Figure \ref{fig:NLQA_Answer}. `\textit{Inglourious Basterds}' is also a candidate, but not selected by the users because there is no ambiguity to solve.

The strengths of this NLI are the simplicity and the efficient user interaction process. The simplicity allows easy adaption on new knowledge graphs and together with the user interaction process it overcomes the difficulties of ambiguity.

The weakness of this system is that usually more than one user interaction is needed to resolve ambiguities, in the experiments the average number of interactions was three \citep{NLQ/A2017}.

\subsubsection{QuestIO (QUESTion-based Interface to Ontologies)}
QuestIO \citep{QuestIO2008} is an NLI to query ontologies using unconstrained natural language. It automatically extracts human-understandable lexicalization from the ontology. Therefore, the quality of the semantic information in the ontology has to be very high to contain enough human-understandable labels and/or descriptions. For example, the attribute \texttt{Movie.Release}-\texttt{Date} would be extracted as `\textit{Release Date}', which is a human-understandable label. In contrast, the attribute \texttt{Movie.OriginalLang} would result in `\textit{Original Lang}', where the token `\textit{Lang}' is a shortened version for `\textit{Language}' and is not human-understandable.

QuestIO translates the input question with three steps: In the first step, the \textit{key concept identification tool} identifies all tokens which refer to mentions of ontology resources such as instances, classes, properties or property values. This is similar to the dependent phrases of NLQ/A. In the next step, the context collector identifies patterns (e.g., key phrases like `\textit{how many}') in the remaining tokens that help the system to understand the query (similar to independent phrases of NLQ/A). The last step identifies relationships between the ontology resources collected during the previous steps and formulates the corresponding formal query. After executing the query, it will be sent to the \textit{result formatter} to display the result in an user-friendly manner.

The automatic extraction of semantic information out of the ontology is both a strength and a weakness of QuestIO. It is highly dependent on the development of the human-understandable labels and descriptions, without them QuestIO will not be able to match the input questions to the automatic extracted information.

\subsection{Parsing-based systems}
\label{sec:parsing-based}
Parsing-based NLIs are going a step further than previously discussed systems: they parse the input question and use the information generated about the structure of the question to understand the grammatical structure. For example, the grammatical structure can be used to identify the dependencies given by the trigger word `\textit{by}' in a question. This is needed for long range dependencies which cannot be caught with simple natural language patterns. Furthermore, the dependency parser can help to handle the difficulty of verboseness. For example, the nominal modifier (\texttt{nmod}) could be used to identify aggregations.

In the following, we summarize eight parsing-based NLIs. We decided to describe ATHENA \citep{Athena2016} in depth, because it can answer the most of the sample input questions. Furthermore, ATHENA uses the most NLP technologies and the authors describe all the steps in depth. Afterwards, the other systems are summarized and we highlight the delta to ATHENA and previous systems.

\subsubsection{ATHENA}
ATHENA \citep{Athena2016} is an ontology-driven NLI for relational databases, which handles full sentences in English as the input question. For ATHENA, ontology-driven means that it is based on the information of a given ontology and needs mapping between an ontology and a relational database. A set of synonyms can be associated with each ontology element. During the translation of an input question into a SQL query, ATHENA uses an intermediate query language before subsequently translating it into SQL. 
 
Assuming the users want to know the director of the movie `\textit{Inglourious Basterds}' (\texttt{Q1}), the input question for ATHENA could be: `\textit{Who is the director of ``Inglourious Basterds''?}'.

ATHENA uses four steps to translate a full sentence input question into a SQL query. In the first step, the ontology evidence annotator is used, which maps the input to a set of ontology elements. There are five types of possible matches:

\begin{enumerate}[a.]
\item \textit{meta data}: Finding a match in the inverted index for the meta data (and the associated set of synonyms). Longer matches over the input question are preferred if there are multiple matches.

\item \textit{translation index}: The translation index is an extension of the inverted index over the base data, which is enriched with variations for person and company names. For example, for the person name `\textit{Brad Pitt}' there would also be an entry `\textit{B. Pitt}'. 

\item \textit{time range expressions}: Finding all time ranges like `\textit{from 2000 until 2010}' (\texttt{Q5}) with the TIMEX annotator. Those time ranges are then matched to the ontology properties with the corresponding data type.

\item \textit{numeric expressions}: Finding all tokens that include numeric quantities with the Stanford Numeric Expressions annotator. The numeric quantities can be either in the form of numbers (e.g., 9) or in text form (e.g., nine). Those numeric expressions are then matched to the ontology properties with the corresponding data type.

\item \textit{dependencies}: Annotating dependencies between tokens in the input question. For example, in the input question \texttt{Q1} there is a dependency between the tokens `\textit{director}' and `\textit{Inglourious Basterds}' indicated by the token `\textit{of}'.
\end{enumerate}

\begin{figure}
	\includegraphics[width=\columnwidth]{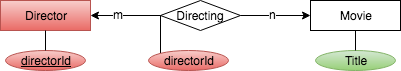}
	\caption{Matches found by ATHENA with the ontology evidence annotator for the input question `\textit{Who is the director of ``Inglourious Basterds''?}'}
	\label{fig:ATHENA_step1}
\end{figure}

For the input question \texttt{Q1} the meta data annotation will detect three different matches for `\textit{director}', namely the table name \texttt{Director} and the attribute name \texttt{Director.directorId} and \texttt{Directing.director}-\texttt{Id} (Figure \ref{fig:ATHENA_step1}: red). The translation index will find a match for the bi-gram `\textit{Inglourious Basterds}', corresponding to the attribute \texttt{Movie.Title} (Figure \ref{fig:ATHENA_step1}: green).

The next step generates a ranked list of interpretations. An interpretation is a set of ontology elements provided by the previous step. If n ontology elements exist for one token, there will also be n different interpretations, one for each ontology element. For the given input question there are three different interpretations possible: \{\texttt{Directing.directorId}, \texttt{Movie.Title}\}, \linebreak
\{\texttt{Director.directorId}, \texttt{Movie.Title}\} and \{\texttt{Director}, \texttt{Movie.Title}\}. Each interpretation is represented by a set of interpretation trees. An \textit{interpretation tree} (iTree) is a subtree of the ontology. Each iTree must satisfy:

\begin{enumerate}[a.]
\item \textit{evidence cover}: All tokens, which were annotated in the previous step, need to be covered. 

\item \textit{weak connectedness}: All concepts need to be at least weakly connected through an undirected path and each property must be connected to its corresponding concept. For the first interpretation this means that \texttt{Director} and \texttt{Movie} need to be connected, for example, via the relation \texttt{Directing}. The attribute \texttt{Title} needs to be connected with the corresponding concept (in this case the table) \texttt{Movie}.

\item \textit{inheritance constraint}: No ontology element is allowed to inherit from its child concepts. For example, the ontology element \texttt{Person} is not allowed to inherit \texttt{Role} of \texttt{Actor}. The other direction is allowed, such that \texttt{Actor} inherits \texttt{FirstName} and \texttt{LastName} from \texttt{Person}.

\item \textit{relationship constraint}: All relationships given in the input question are included, not depending on the direction of the path. For example, the tree tokens `\textit{movies}', `\textit{starring}' and `\textit{Brad Pitt}' (\texttt{Q5}) imply a relationship constraint between the ontology element \texttt{Movie}, \texttt{Starring} and \texttt{Person}. Those three ontology elements need to be connected. Accordingly, in this example the ontology element \texttt{Actor} needs to be included.
\end{enumerate}

\begin{figure}
	\includegraphics[width=\columnwidth]{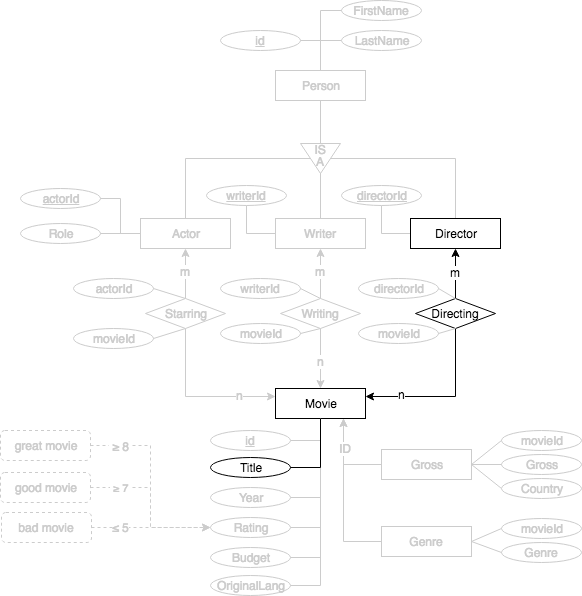}
	\caption{Possible interpretation tree (black) for the input question `\textit{Who is the director of ``Inglourious Basterds''?}' (\texttt{Q1})}
	\label{fig:ATHENA_iTree}
\end{figure}

For the purpose of ranking the different interpretations, ATHENA generates one single iTree. It can consist of a union of multiple iTrees or a single iTree. Figure \ref{fig:ATHENA_iTree} shows a possible iTree for the interpretation \{\texttt{Director}, \texttt{Movie.Title}\}, which is extended with the ontology element \texttt{Directing} and \texttt{Movie}. After this step, for each interpretation only one iTree is left.

The third step uses the ranked list of interpretations to generate an intermediate query in the \textit{Ontology Query Language} (OQL). OQL was specifically developed for ATHENA to be an intermediate language between the input question and SQL and is able to express queries that include aggregations, unions and single nested subqueries. The structure of an OQL query is similar to SQL and is generated as follows:

\begin{enumerate}[a.]
\item \texttt{from} \textit{clause}: Specifies all concepts found in the ontology along with their aliases. The aliases are needed, for example, if a concept occurs multiple times. For example, the input question `\textit{Show me all drama and comedy movies.}' (\texttt{Q4}) would point to \texttt{Genre} in the ontology twice: once for the token `\textit{drama}' and once for `\textit{comedy}'. Therefore, two aliases are needed to distinguish between them.

\item \texttt{group by} \textit{clause}: The \texttt{group by} clause is triggered by the word `\textit{by}'  and only tokens annotated with meta data in step 1.a are considered. For example, the input question `\textit{What was the best movie by genre?}' (modified \texttt{Q7}). To identify the dependencies between dependent and dependee (illustrated by the `\textit{by}'), the Stanford Dependency Parser is used.

\item \texttt{select} \textit{clause}: There are two possible types: \textit{aggregation} and \textit{display} properties. The \textit{aggregation} properties depend on the \texttt{group by} clause. The default aggregation function is \texttt{sum}. For the (modified) input question \texttt{Q7}, ATHENA would detect a \texttt{group by} clause because the `\textit{by genre}' needs an aggregation function. Assuming ATHENA can translate `\textit{best movie}' to mean `\textit{best ranked movie}', it would apply the aggregation function \texttt{max} on \texttt{Movie.Rating}. If there are no aggregations, ATHENA uses the tokens annotated with meta data as \textit{display} properties, which are shown to the user. 

\item \texttt{order by} \textit{clause}: Properties used in the \texttt{order by} clause are indicated by tokens like `\textit{least}', `\textit{most}', `\textit{ordered by}', `\textit{top}' and others. For example, the input question `\textit{Which movie has grossed most?}' (\texttt{Q3}) would trigger an \texttt{order by} clause for \texttt{Movie.Gross} because of the trigger word `\textit{most}'.

\item \texttt{where} \textit{clause}: Tokens annotated with the translation index, time range or numerical expression are used in the \texttt{where} clause to filter the result (e.g., the tokens `\textit{Inglourious Basterds}'). If the filter is applied on an aggregation, a \texttt{having} clause is generated instead of the \texttt{where} clause.
\end{enumerate}

The final step translates the OQL query into a SQL query, where each attribute and join condition is either \textit{concrete} or \textit{virtual}. \textit{Concrete} means that a direct mapping between ontology and the database (e.g., `\textit{director}') exist. \textit{Virtual} implies that a (complex) relationship between the ontology elements and the database (e.g., `\textit{great movie}') exists. Furthermore, the result of the best ranked interpretation is directly displayed, but the users will see the other interpretations as well. All the top \texttt{n} interpretations that ATHENA has found are translated back into full sentences in English for the users, so that the users can choose the best fitting one.

The strengths of ATHENA are the ontology as an abstraction of the relational database and the natural language explanation for each interpretation of the input question. The translation index contains not only synonyms but also semantic variants for certain types of values like persons and company names. Furthermore, ATHENA can handle single level nesting in the input question. An example input question could be `\textit{All movies with rating higher than the rating of ``Sin City''.}' (\texttt{Q9}).

One of the weaknesses of ATHENA is that neither negation (\texttt{Q8}) nor multiple elements in the \texttt{group by} clause (e.g., `\textit{What was the best movie by year and}\linebreak \textit{genre?}') are supported. 

\cite{Athena2016} suggest extending ATHENA to handle more than single level nesting. Furthermore, they suggest enabling a possibility to answer follow-up questions using the context of the previous questions.

\subsubsection{Querix}
Querix\footnote{The name is based on the druid Getafix who is consulted by Asterix (in the same named comic) whenever anything strange occurs.} \citep{Querix2006} allows users to enter questions in natural language to query an ontology. If the system identifies any ambiguities in the input question, it asks the user for clarification in a dialog. Querix uses a syntax tree to extract the sequence of words from the main word categories: noun (\texttt{N}), verb~(\texttt{V}), preposition (\texttt{P}), \texttt{wh}-pronoun (\texttt{Q}, e.g. what, where, when, etc.) and conjunction (\texttt{C}). This sequence is called query skeleton. The query skeleton is used to enrich nouns and verbs and to identify \texttt{subject}-\texttt{property-object} patterns in the query.

In contrast to ATHENA, which uses a lot of different tools and technologies, Querix only uses the information of the query skeleton (parse tree) and the synonyms (for both the input question and the ontology) to translate the input question into SPARQL. For translating into SPARQL, Querix uses three components: \textit{query analyzer}, \textit{matching center} and \textit{query generator}. The \textit{query analyzer} handles two tasks: (1) It applies the Stanford Parser on the input question to generate a syntax tree, from which Querix extracts the query skeleton. For example, the query skeleton `\texttt{Q-V-N-P-N}' is extracted from the input question (\texttt{Q1}) as `\textit{Who} (\texttt{Q}) \textit{is} (\texttt{V}) \textit{the director} (\texttt{N}) \textit{of} (\texttt{P}) \textit{``Inglourious Basterds''} (\texttt{N}) \textit{?}'. (2) It enriches all nouns and verbs with synonyms provided by WordNet. 

The \textit{matching center} is the core component of Querix: (1) It tries to match the query skeleton with a small set of heuristic patterns. Those patterns are used to basically identify \texttt{subject-property-object} patterns in the input question. (2) It searches for matches between nouns and verbs of the input question with the resources in the ontology (including synonyms). (3) It tries to match the results of the two previous steps. 
The \textit{query generator} then composes SPARQL queries from the joined triplets delivered by the last step of the matching center. If there are several different solutions with the highest cost score, Querix will consult the user by showing a menu from which the user can choose the intended meaning.

The simplicity of Querix is both a strength and a weakness: it is simple to use and completely portable, but this simplicity also reduces the number of questions that can be answered since they have to adhere to a predefined syntax.

\subsubsection{FREyA (Feedback, Refinement and Extended vocabularY Aggregation)}
FREyA \citep{FREyA2010} is based on QuestIO. It allows users to enter queries in any form in English. It generates a syntactic parse tree in order to identify the answer type. The translation process starts with a lookup, annotating query terms with ontology concepts by using heuristic rules. If there are ambiguous annotations, the user will be engaged in a clarification dialogue. The user's selections are saved and used for training the system in order to improve its performance over time. In the end, the system generates a SPARQL query.

The strength of this system is the user interaction, which not only supports the users to find the right answer, but also improves FREyA over time. Furthermore, FREyA performs an answer type identification, which leads to more precise answers. The weakness is that only a few of the questions could be answered without any clarification dialogs. Furthermore, the system cannot answer negations.

\subsubsection{BELA}
BELA \citep{BELA2012} is an NLI with a layered approach. This means, that at each layer the best hypothesis is determined. If the confidence for a particular interpretation is \texttt{1} and the SPARQL query generated by it produces an answer with at least one result, the translation process is stopped and the answer is returned to the user. Only for \texttt{ASK}-questions (which have yes/no answers), the process continues until the confidence of the interpretations start to differ, then a threshold of 0.9 is applied and an empty result (which equals a \textit{no}-answer) is also accepted.

Similar to Querix, BELA parses the input question and produces a set of query templates, which mirror the semantic structure. The next step is a lookup in the index, including Wikipedia redirects (this corresponds to the translation index of ATHENA). The first lookup is without fuzzy matching, if no result can be found, a threshold of 0.95 is applied for a matching with normalized Levenshtein distance. The third lookup enables the usage of synonyms, hypernyms and hyponyms from WordNet. The last lookup (and step) uses \textit{Explicit Semantic Analysis}, which can be used to relate expressions like `\textit{playing}' to concepts like `\textit{actor}'.

Compared to other systems, BELA not only focuses on solving the translation task, but also reduces the computation time, which increases the user-friendliness.

\subsubsection{USI Answers}
USI Answers \citep{USIAnswers2013} is an NLI for semi-structured industry data. It offers natural language access to the large bodies of data that are used in the planning and delivery of services by Siemens Energy. The database consists of multiple databases, for example domain-specific ontologies, different relational databases and SPARQL endpoints. These can be either internal or external knowledge (e.g., DBpedia). Furthermore, the users demanded to be able to use not only natural language questions and formal query language constructs but also keyword questions or a mixture of these.

In contrast to the previous parsing-based NLIs, USI Answers cannot rely on the parse tree, because of the different types of possible input (e.g., keyword questions). Still, the first step includes various NLP technologies, such as lemmatization, PoS tagging, named entity recognition and dependency parsing. To handle domain-specific input terms, there is a list of syntax rules to revalidate the entity information after the NLP technologies were applied. Furthermore, the question focus (i.e., referenced entity object) is identified by applying 12 syntactic rules. Enriched with this information, the next step aims to identify and resolve the different concepts that may be in the input question (lookup). Next, USI Answers detects and prevalidates the different relations and objects found in the previous step. The fourth step generates different question interpretations, including how concepts and instances are connected to each other. Afterwards, the different interpretations are validated and ranked. This is done with a learned model, based on user-defined views. The last step constructs the final query in the representative query language.

One strength of USI Answers is the ability to query different database resources. Furthermore, the users are free to choose their preferred type of input questions.

\subsubsection{NaLIX (Natural Language Interface to XML)}
NaLIX \citep{NaLIX2005} is an interactive NLI for querying an XML database with XQuery. The interaction is based on guiding the users to pose questions that the system can handle by providing meaningful feedback and helpful rephrasing suggestions. Furthermore,  NaLIX provides templates and question history to the users. It preserves the users prior search efforts and provides the users with a quick starting point when they create new questions.

Similar to Querix, NaLIX mostly uses the parse tree for the translation of an input question into XQuery. After MiniPar\footnote{\url{https://gate.ac.uk/releases/gate-7.0-build4195-ALL/doc/tao/splitch17.html}} is used to parse the input question, NaLIX identifies the phrases in the parse tree of the input question that can be mapped to XQuery components and classifies them. In the next step, NaLIX validates the classified parse tree from the previous step: it checks if it knows how to translate the classified parse tree into XQuery and if all attribute names and values can be found in the database. The last step translates the classified parse tree into an appropriate XQuery expression if possible. During both the classification and the validation of the parse tree, information about errors (e.g.,~unknown phrases and invalid parse trees) is collected to report to the users for clarification.

The strength of NaLIX is the ability to solve difficult questions, which can include subqueries, by using and adjusting the parse tree. On the other hand, the reliance on a parse tree is also a weakness, because the system can only answer questions that are parseable.

\subsubsection{NaLIR (Natural Language Interface for Relational databases)}
NaLIR \citep{NaLIR2014, NaLIR2014a} is a further development of NaLIX. Instead of translating into XQuery to query XML databases, NaLIR translates the input question into SQL and queries relational databases.\linebreak NaLIR returns not only the result to the user, but also a rephrased version of the input question based on the translation.

The basic idea remains the same: parse the input question using the Stanford Parser and map the parse tree to SQL (instead of XQuery). The steps are similar with some modifications: In the step of mapping phrases of the input question parse tree to SQL components, the users are asked for clarification if there are ambiguities. The next step is not a validation anymore, but an adjustment of the parse tree in such a way, that it is valid. The candidate interpretations produced by this adjusted and valid parse tree are delivered to the users to select the correct interpretation. The last step translates the adjusted and valid parse tree which the user has selected into SQL.

The strength of NaLIR is the interaction with the user, which improved further compared to NaLIX. The weakness remains: it is highly dependent on the parse tree.

\subsubsection{BioSmart}
BioSmart \citep{BioSmart2017} uses a syntactic classification of the input question to translate the natural language question into a declarative language such as SQL. The system divides the input questions into three query types (iterative, conditional or imperative) using the parse tree of the input question and compare it to parse tree templates for each query type. For example, a imperative query consists of a \textit{verb} (\texttt{VB}) followed by a \textit{object} (\texttt{NP}). A more expressive and therefore complex input question can be built by nesting simple query types arbitrarily.

Similar to Querix, BioSmart uses the Stanford parser to parse the input question. The system then tries to map the resulting parse tree to predefined questions or to one of the query templates (query type identification). As mentioned, it is possible that a question consists of several of these templates to capture the meaning of the question. Then the tables and possible joins that are needed to compute the query are identified. Afterwards, the templates are transformed into a logical query using the information about the tables and joins. 

Compared to other NLI, the strength of BioSmart is the possibility to query arbitrary databases. The weakness of BioSmart is the mapping to the three query types: if the system cannot match the input question to those query types, it is not able to answer the question.

\subsection{Grammar-based systems}
\label{sec:grammar-based}
Grammar-based NLIs use a different approach. The core of these systems is a set of rules (grammar) that defines the questions that can be understood and answered by the system. Using those rules, the system is able to give the users suggestions on how to complete their questions during typing. This supports the users to write understandable questions and gives the users insight into the type of questions that can be asked. The disadvantage of these systems is that they are highly domain dependent: the rules need to be written by hand.

In the following, we summarize seven grammar-based NLIs. We decided to describe TR Discover \citep{TRDiscover2015} in depth, because it is well described in the paper such that we can provide examples for the whole process, including the grammar rules. Furthermore, it uses the rules to guide and help the users during formulating their questions. Afterwards, other grammar-based systems are summarized and we describe the delta to TR Discover. The differences can be quite significant, however, they all have the same core: a set of rules.

\subsubsection{TR Discover}
TR Discover \citep{TRDiscover2015} is a system providing an NLI which translates the input question in form of an English sentence (or sentence fragment) into SQL or SPARQL. It is either used for relational databases or ontologies, but does not need an ontology to work for relational databases. During the translation steps, TR Discover uses a \textit{First Order Logic} (FOL) representation as an intermediate language. Furthermore, it provides auto-suggestions based on the user input. There are two types of suggestions: autocompletion and prediction.

TR Discover helps the users in formulating the question through an auto-suggestion feature. For example, assuming the users want to know the director of the movie `\textit{Inglourious Basterds}' (\texttt{Q1}). When the users start typing `\textit{p}', TR Discover will not only suggest `\textit{person}' but also longer phrases like `\textit{person directing}' (autocomplete). After `\textit{person directing}' is selected (or typed), TR Discover will again suggest phrases, like `\textit{movies}' or even specific movies like `\textit{Inglourious Basterds}' (prediction). For input question \texttt{Q1}, the input could be `\textit{person directing Inglourious Basterds}'.

The suggestions are based upon the relationships and entities in the dataset and use the linguistic constraints encoded in a \textit{feature-based context-free grammar} (FCFG). The grammar consists of grammar rules (\texttt{G1-3}) and lexical entries (\texttt{L1-2}). For the sample world (and the input question \texttt{Q1}), the following rules could be defined:

\begin{description}
  \item[\texttt{G1}:] \texttt{NP $\rightarrow$ N}
  \item[\texttt{G2}:] \texttt{NP $\rightarrow$ NP VP}
  \item[\texttt{G3}:] \texttt{VP $\rightarrow$ V NP}
  \item[\texttt{}{L1}:] \texttt{N[TYPE=person, NUM=sg, SEM=<x.person(x)>] \\$\rightarrow$ person}
  \item[\texttt{L2}:] \texttt{V[TYPE=[person,movie,title], \\SEM=<X x.X(y.directMovie(y,x)>, TNS=presp] \\$\rightarrow$ directing}
\end{description}

The suggestions are computed based on the idea of left-corner parsing: given a query segment, it finds all grammar rules whose left corner on the right side matches the left side of the lexical entry of the query segment. Then, all leaf nodes (lexical entries) in the grammar that can be reached by using the adjacent element are found. For example, while typing `\textit{person}' (\texttt{Q1}), the lexical entries \texttt{L1} and \texttt{L2} are found and provided to the user.

TR Discover uses three steps to translate the English sentence or fragment of a sentence into a SQL or SPARQL query. The first step parses the input question into a FOL representation. The query parsing uses the FCFG. For the example input, the token `\textit{person}' will be parsed by the lexical entry \texttt{L1} and the token `\textit{directing}' will be parsed with the lexical entry \texttt{L2}. This leads to the FOL representation: \newline
\texttt{x.person(x) $\rightarrow$ directMovie(y,x)\& type(y,Movie) \& label(y, `Inglourious Basterds')}

How exactly the phrase `\textit{Inglourious Basterds}' is matched to the base data and therefore can be used as part of the lexical entry \texttt{L2} and how it is resolved, is not explained by \cite{TRDiscover2015}. If there are multiple possibilities to parse the input question, the first one is chosen.

\begin{figure}
	\includegraphics[width=\columnwidth]{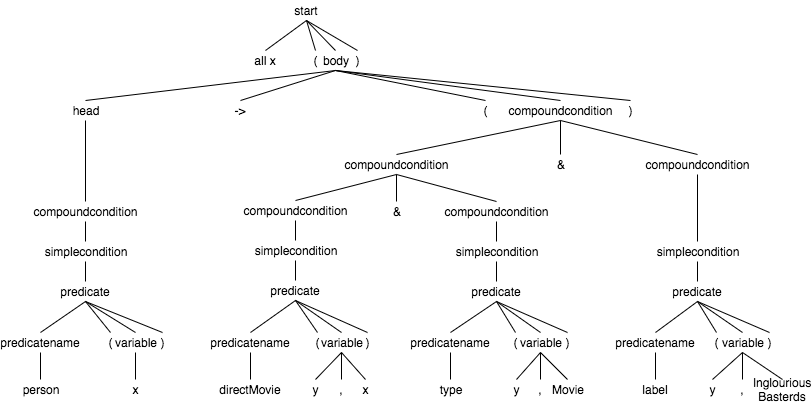}
	\caption{The parse tree for the FOL representation of the input question `\textit{person directing Inglourious Basterds}'.}
	\label{fig:TRDiscover_parseTree}
\end{figure}

The next step is to translate the generated FOL into a parse tree. The FOL parser takes a grammar and the FOL representation from the previous step, and generates a parse tree (Figure \ref{fig:TRDiscover_parseTree}) using ANTLER for implementation.

In the third step, an in-order traversal of the parse tree (provided by the previous step) is performed to translate it into an executable SQL or SPARQL query. While traversing the parse tree, the atomic logical conditions and connectors are put on a stack. After traversing, the constraints are popped from the stack to build the correct query constraints. The predicates are mapped to their corresponding attribute names (SQL) or ontology properties (SPARQL).

The strengths of TR Discover are the auto-suggestion and the possibility to translate natural language into different query languages such as SQL and SPARQL, because FOL is used as an intermediate language.

The weaknesses of TR Discover are that quantifiers (e.g., \texttt{Q3}: `\textit{grossed most}') cannot be used, synonyms are not properly handled, and negations only work for SPARQL.

\cite{TRDiscover2015} suggest extending TR Discover with a ranking system for multiple parses in the first step and to improve the synonym handling. Furthermore, they pointed out the possibility of applying user query logs to improve auto-suggestions.

\subsubsection{Ginseng (Guided Input Natural language Search ENGine)}
Ginseng \citep{Ginseng2005} is a guided input NLI for ontologies. The system is based on a grammar that describes both the parse rules of the input questions and the query composition elements for the \textit{RDF Data Query Language} (RDQL) queries. The grammar is used to guide the users in formulating questions in English.

In contrast to TR Discover, Ginseng does not use an intermediate representation and therefore the parsing process translates directly into RDQL. The grammar rules are divided in two categories: \textit{dynamic} and \textit{static} grammar rules. The \textit{dynamic} grammar rules are generated from the OWL ontologies. They include rules for each class, instance, objects property, data type property and synonyms. The \textit{static} grammar rules consist of about 120 mostly empirically constructed domain-independent rules, which provide the basic sentence structures and phrases for input questions. The naming conventions used by Ginseng differ slightly from these used by TR Discover. Ginseng's dynamic rules correspond to TR Discover's lexical rules and Ginseng's static rules consists of both grammar and lexical rules in TR Discover. 

The strength of Ginseng are the dynamic rules which are generated from the ontology. This, together with the domain-independent static rules, lead to an easier adaptability compared to systems like TR Discover. The weakness lies in the grammar rules: they need to cover all possible types of questions the users want to ask.

\subsubsection{SQUALL (Semantic Query and Update High-Level Language)}
SQUALL \citep{SQUALL2014,SQUALL2011} is an NLI for searching and updating an RDF store. It uses the style of Montague grammars (context-free generative grammar) as an intermediate language (similar to TR Discover) to split the translation process in two parts: translating the natural language input question into a logical language and translating the logical language into a query language. Because of that, the second part is getting easier: the logical language and the query language share the same semantics and level of detail. The grammar of SQUALL consists of about 120 domain-independent rules.

The translation into the logical form is done in three steps. In the first step, the keywords are recognized (lookup step). The second step is a syntactic analysis based on a descending parser, which is fed with the grammar rules. Afterwards, the next step can generate the logical language based on the definition in the grammar. After the translation into the logical language, the translation in to the chosen formal language can be done.

The strength of SQUALL is that it is able to translate any type of input question, including aggregations, negations and subqueries. The weakness of SQUALL is that the users have to know the RDF vocabulary (e.g.,~classes and properties). For example, the input question \texttt{Q1} needs to be formulated as `\textit{Who is the director of \texttt{Inglourious\_Basterds}?}'.

\subsubsection{MEANS (MEdical question ANSwering)}
MEANS \citep{MEANS2015} is an NLI that uses a hybrid approach of patterns and ML to identify semantic relations. It is highly domain dependent and focuses on factual questions expressed by \texttt{wh}-pronouns and boolean questions in a medical subfield targeting the seven medical categories: problem, treatment, test, sign/symptom, drug, food and patient.

To translate the input question into SPARQL,\linebreak MEANS first classifies the input question into one of ten categories (e.g., factoid, list, definition, etc.). If the question is categorized as a \texttt{wh}-question, the \textit{Expected Answer Type} (EAT) is identified and replaced with\linebreak `\texttt{ANSWER}' as a simplified form for the next step. For example, the EAT of the input question \texttt{Q1} would be `\textit{director}'. In the next step, MEANS identifies medical entities using a \textit{Conditional Random Field} (CRF) classifier and rules to map noun phrases to concepts. The next step is used to identify seven predefined semantic relations. The annotator is a hybrid approach based on a set of manually constructed patterns and a \textit{Support Vector Machine} (SVM) classifier.

The strength of MEANS is, that it can handle different types of questions, including questions with more than one expected answer type and more than one focus. As for most of the grammar-based NLIs, MEANS suffers from the restriction based on the handcrafted rules. The inclusion of ML reduces this problem, but ML itself needs huge training corpus to be usable. Furthermore, comparison (and also negation) are not taken into account.

\subsubsection{AskNow}
AskNow \citep{AskNow2016} uses a novel query characterization structure that is resilient to paraphrasing, called \textit{Normalized Query Structure} (NQS), which is less sensitive to structural variation in the input question. The identification of the elements in the NQS is highly dependent on POS tags. For example, the input question \texttt{Q1} `\textit{Who is the director of ``Inglourious Basterds''?}' would be matched to the NQS template: \\\texttt{[Wh][R1][D][R2][I]}, where \texttt{[Wh]} is the question word `\textit{Who}', \texttt{[R1]} is the auxiliary relation `\textit{is}', \texttt{[D]} is the query desire class `\textit{director}', \texttt{[R2]} the relation `\textit{of}' and \texttt{[I]} is the query input class `\textit{Inglourious Basterds}'.

To translate the input question into SPARQL, AskNow first identifies the sub-structures using a POS tagger and named entity recognition. Then it fits the sub-structures into their corresponding cells within the generic NQS templates. Afterwards the query type (set, boolean, ranking, count or property value) is identified based on desire and \texttt{wh}-type. In the next step, the query desire, query input and their relations will be matched to the KB. As an example, Spotlight can be used for the matching to DBpedia. During the matching process, AskNow uses WordNet synonyms and a BOA pattern library (bootstrapping).

The strength of AskNow, compared to the previous grammar-based systems, is that the users are free to formulate their questions without restrictions. In addition, the NQS templates allow complex questions which, for example, can include subqueries. One weakness of AskNow is that it highly depends on the right PoS tags and restricts the types of questions that can be asked.

\subsubsection{SPARKLIS}
SPARKLIS \citep{SPARKLIS2017} is a guided query builder for SPARQL using natural language for better understanding. It guides the users during their query phrasing by giving the possibilities to search through concepts, entities and modifiers in natural language. It relies on the rules of SPARQL to ensure syntactically correct SPARQL queries all the time during the process. The interaction with the system makes the question formulation more constrained, slower and less spontaneous, but it provides guidance and safeness with intermediate answers and suggestions at each step. The translation process for SPARKLIS is reversed: it translates possible SPARQL queries into natural language such that the users can understand their choices.

The auto-completion is different from the previous systems (e.g., TR Discover and Ginseng): the interface displays three lists where the users can search for concepts, entities or modifiers. To ensure completeness relative to user input, SPARKLIS uses a cascade of three stages. The first stage is on client side, where a partial list of suggestion is filtered. The second stage is executed if the filtered list gets empty, then the suggestions is re-computed by sending the query, including the users filter, to the SPARQL endpoint. The last stage is triggered if the list is still empty, then, new queries are again sent to the SPARQL endpoint, using the full SPARQL query instead of partial results. Only a limited number of suggestions are computed and no ranking is applied, because of scalability issues. 

The strength of SPARKLIS is also its weakness: the restricted guidance of the users during the query formulation process allows only syntactically correct questions, but at the same time, the users' freedom is reduced. Furthermore, the limited number of suggestions has the negative consequence that they may be incomplete and therefore making some queries unreachable.

\subsubsection{GFMed}
GFMed \citep{GFMed2017} is an NLI for biomedical linked data. It applies grammars manually built with \textit{Grammatical Framework}\footnote{\url{https://www.grammaticalframework.org/}} (GF). GF grammars are divided into \textit{abstract} and \textit{concrete} grammars. The \textit{abstract} grammar defines the semantic model of the input language and for GFMed this is based on the biomedical domain. The \textit{concrete} grammars define the syntax of the input language, which is English and SPARQL. Furthermore, GF supports multilingual applications and because of that Romanian is included as a second natural language in GFMed. 

To translate the controlled natural language input into SPARQL, GFMed relies in a first step on the libraries of GF for syntax, morphological paradigms and coordination. GFMed covers basic elements of SPARQL to support term constraints, aggregates and negations. There is no support for property paths of length different from 1, optional graph pattern or assignment. Furthermore, only equality and regular expression operators are included.

The strength of GFMed is that it covers the basic elements of SPARQL. Furthermore, it introduces a second natural language besides of English for the users to ask questions. The weakness are the GF grammars which are domain dependent and restrict the number of questions that can be asked by the users.

\section{Evaluation}
\label{sec:NLI-Eval}
In this section, we first evaluate the 24 recently developed NLIs which were summarized before and systematically analyze if they are able to handle the ten sample questions of increasing complexity. This approach enables us to directly compare these systems which was previously not possible due to different datasets or evaluation measures used in the original papers. Afterwards, we evaluate three commercial systems by asking them the same sample questions and analyzing their responses.

\subsection{Evaluation of 24 recently developed NLIs}
\label{sec:NLI-summary}
In this section, we provide a systematic analysis of the major NLIs. We categorized and analyzed the translation process of the 24 recently developed systems highlighted in Section \ref{sec:systems} based on ten sample world questions with increasing complexity. Each system category, based on its technical approach, has its own strengths and weaknesses. There are also different aspects on which a system can focus (e.g., number of user interaction, ambiguity, efficiency, etc.), which we do not take into account in this paper.

We evaluate the systems based on what is reported in the papers. If there is either an example of a similar question (e.g. '\textit{Who is the president of the united states}?' (\texttt{Q1}) or a clear statement written in the paper (e.g., '\textit{we can identify aggregations}' (\texttt{Q7}), we label those questions for the system with a checkmark (\ding{51}) in Table \ref{tab:questions}. If the question needs to be asked in a strict syntactical way (e.g., SODA needs the symbol '\textit{>}' instead of '\textit{higher than}') or the answer is partially correct (e.g., \texttt{Q4} returns a ordered list of movies instead of only one), it is labeled with a triangle (\ding{115}). If there is a clear statement that something is not implemented (e.g. ATHENA does not support negations), we label it with \ding{55}. If we were not able to conclude, if a system can or cannot answer a question based on the paper, we labeled it with a question mark in Table \ref{tab:questions}.

\begin{table*}[ht!!]
\centering
\caption{Analysis of recently developed NLIs based on ten sample input questions. (\ding{51}: can answer; \ding{115}: strict syntax or partly answerable; \ding{55}: cannot answer; ?: not documented)}
\label{tab:questions}
\begin{tabular}{lll|cccccccccc|ll}
& & & \textbf{Q1} & \textbf{Q2} & \textbf{Q3} & \textbf{Q4} & \textbf{Q5} & \textbf{Q6} & \textbf{Q7} & \textbf{Q8} & \textbf{Q9} & \multicolumn{1}{l|}{\textbf{Q10}} & \begin{turn}{90}\textbf{SQL}\end{turn} & \begin{turn}{90}\textbf{SPARQL}\end{turn} \\ 
\hline
\multirow{7}{*}{Keyword} & \textbf{SODA} & \textit{2012} & \cellcolor[HTML]{009901}{\color[HTML]{FFFFFF} \ding{51}} &  \cellcolor[HTML]{F8A102}{\color[HTML]{FFFFFF} \ding{115}} & \cellcolor[HTML]{F8A102}{\color[HTML]{FFFFFF} \ding{115}} & \cellcolor[HTML]{CB0000}{\color[HTML]{FFFFFF} \ding{55}} & \cellcolor[HTML]{F8A102}{\color[HTML]{FFFFFF} \ding{115}} & \cellcolor[HTML]{009901}{\color[HTML]{FFFFFF} \ding{51}} & \cellcolor[HTML]{CB0000}{\color[HTML]{FFFFFF} \ding{55}} & \cellcolor[HTML]{CB0000}{\color[HTML]{FFFFFF} \ding{55}} & \cellcolor[HTML]{CB0000}{\color[HTML]{FFFFFF} \ding{55}} & \cellcolor[HTML]{CB0000}{\color[HTML]{FFFFFF} \ding{55}} & \cellcolor[HTML]{009901}{\color[HTML]{FFFFFF} \ding{51}} & \cellcolor[HTML]{CB0000}{\color[HTML]{FFFFFF} \ding{55}} \\
&\textbf{NLP-Reduce} & \textit{2007} & \cellcolor[HTML]{009901}{\color[HTML]{FFFFFF} \ding{51}} & \cellcolor[HTML]{CB0000}{\color[HTML]{FFFFFF} \ding{55}} & \cellcolor[HTML]{C0C0C0}{\color[HTML]{FFFFFF} \textbf{?}} & \cellcolor[HTML]{CB0000}{\color[HTML]{FFFFFF} \ding{55}} & \cellcolor[HTML]{C0C0C0}{\color[HTML]{FFFFFF} \textbf{?}} & \cellcolor[HTML]{009901}{\color[HTML]{FFFFFF} \ding{51}} & \cellcolor[HTML]{CB0000}{\color[HTML]{FFFFFF} \ding{55}} & \cellcolor[HTML]{C0C0C0}{\color[HTML]{FFFFFF} \textbf{?}} & \cellcolor[HTML]{CB0000}{\color[HTML]{FFFFFF} \ding{55}} & \cellcolor[HTML]{CB0000}{\color[HTML]{FFFFFF} \ding{55}} & \cellcolor[HTML]{CB0000}{\color[HTML]{FFFFFF} \ding{55}} & \cellcolor[HTML]{009901}{\color[HTML]{FFFFFF} \ding{51}} \\
&\textbf{Précis} & \textit{2008} & \cellcolor[HTML]{F8A102}{\color[HTML]{FFFFFF} \ding{115}} & \cellcolor[HTML]{CB0000}{\color[HTML]{FFFFFF} \ding{55}} & \cellcolor[HTML]{CB0000}{\color[HTML]{FFFFFF} \ding{55}} & \cellcolor[HTML]{CB0000}{\color[HTML]{FFFFFF} \ding{55}} & \cellcolor[HTML]{F8A102}{\color[HTML]{FFFFFF} \ding{115}} & \cellcolor[HTML]{CB0000}{\color[HTML]{FFFFFF} \ding{55}} & \cellcolor[HTML]{CB0000}{\color[HTML]{FFFFFF} \ding{55}} & \cellcolor[HTML]{F8A102}{\color[HTML]{FFFFFF} \ding{115}} & \cellcolor[HTML]{CB0000}{\color[HTML]{FFFFFF} \ding{55}} & \cellcolor[HTML]{CB0000}{\color[HTML]{FFFFFF} \ding{55}} & \cellcolor[HTML]{009901}{\color[HTML]{FFFFFF} \ding{51}} & \cellcolor[HTML]{CB0000}{\color[HTML]{FFFFFF} \ding{55}} \\
&\textbf{QUICK} & \textit{2009} & \cellcolor[HTML]{009901}{\color[HTML]{FFFFFF} \ding{51}} & \cellcolor[HTML]{CB0000}{\color[HTML]{FFFFFF} \ding{55}} & \cellcolor[HTML]{CB0000}{\color[HTML]{FFFFFF} \ding{55}} & \cellcolor[HTML]{CB0000}{\color[HTML]{FFFFFF} \ding{55}} & \cellcolor[HTML]{CB0000}{\color[HTML]{FFFFFF} \ding{55}} & \cellcolor[HTML]{C0C0C0}{\color[HTML]{FFFFFF} \textbf{?}} & \cellcolor[HTML]{CB0000}{\color[HTML]{FFFFFF} \ding{55}} & \cellcolor[HTML]{C0C0C0}{\color[HTML]{FFFFFF} \textbf{?}} & \cellcolor[HTML]{CB0000}{\color[HTML]{FFFFFF} \ding{55}} & \cellcolor[HTML]{CB0000}{\color[HTML]{FFFFFF} \ding{55}} & \cellcolor[HTML]{CB0000}{\color[HTML]{FFFFFF} \ding{55}} & \cellcolor[HTML]{009901}{\color[HTML]{FFFFFF} \ding{51}} \\
&\textbf{QUEST} & \textit{2013} & \cellcolor[HTML]{009901}{\color[HTML]{FFFFFF} \ding{51}} & \cellcolor[HTML]{C0C0C0}{\color[HTML]{FFFFFF} \textbf{?}} & \cellcolor[HTML]{C0C0C0}{\color[HTML]{FFFFFF} \textbf{?}} & \cellcolor[HTML]{C0C0C0}{\color[HTML]{FFFFFF} \textbf{?}} & \cellcolor[HTML]{C0C0C0}{\color[HTML]{FFFFFF} \textbf{?}} & \cellcolor[HTML]{CB0000}{\color[HTML]{FFFFFF} \ding{55}} & \cellcolor[HTML]{C0C0C0}{\color[HTML]{FFFFFF} \textbf{?}} & \cellcolor[HTML]{C0C0C0}{\color[HTML]{FFFFFF} \textbf{?}} & \cellcolor[HTML]{CB0000}{\color[HTML]{FFFFFF} \ding{55}} & \cellcolor[HTML]{CB0000}{\color[HTML]{FFFFFF} \ding{55}}  & \cellcolor[HTML]{009901}{\color[HTML]{FFFFFF} \ding{51}} & \cellcolor[HTML]{CB0000}{\color[HTML]{FFFFFF} \ding{55}}\\
&\textbf{SINA} & \textit{2015} & \cellcolor[HTML]{009901}{\color[HTML]{FFFFFF} \ding{51}} & \cellcolor[HTML]{C0C0C0}{\color[HTML]{FFFFFF} \textbf{?}} & \cellcolor[HTML]{C0C0C0}{\color[HTML]{FFFFFF} \textbf{?}} & \cellcolor[HTML]{CB0000}{\color[HTML]{FFFFFF} \ding{55}} & \cellcolor[HTML]{F8A102}{\color[HTML]{FFFFFF} \ding{115}} & \cellcolor[HTML]{CB0000}{\color[HTML]{FFFFFF} \ding{55}} & \cellcolor[HTML]{CB0000}{\color[HTML]{FFFFFF} \ding{55}} & \cellcolor[HTML]{C0C0C0}{\color[HTML]{FFFFFF} \textbf{?}} & \cellcolor[HTML]{CB0000}{\color[HTML]{FFFFFF} \ding{55}} & \cellcolor[HTML]{CB0000}{\color[HTML]{FFFFFF} \ding{55}} & \cellcolor[HTML]{CB0000}{\color[HTML]{FFFFFF} \ding{55}} & \cellcolor[HTML]{009901}{\color[HTML]{FFFFFF} \ding{51}} \\
&\textbf{Aqqu} & \textit{2015} & \cellcolor[HTML]{009901}{\color[HTML]{FFFFFF} \ding{51}} & \cellcolor[HTML]{C0C0C0}{\color[HTML]{FFFFFF} \textbf{?}} & \cellcolor[HTML]{C0C0C0}{\color[HTML]{FFFFFF} \textbf{?}} & \cellcolor[HTML]{C0C0C0}{\color[HTML]{FFFFFF} \textbf{?}} & \cellcolor[HTML]{C0C0C0}{\color[HTML]{FFFFFF} \textbf{?}} & \cellcolor[HTML]{C0C0C0}{\color[HTML]{FFFFFF} \textbf{?}} & \cellcolor[HTML]{CB0000}{\color[HTML]{FFFFFF} \ding{55}} & \cellcolor[HTML]{C0C0C0}{\color[HTML]{FFFFFF} \textbf{?}} & \cellcolor[HTML]{CB0000}{\color[HTML]{FFFFFF} \ding{55}} & \cellcolor[HTML]{CB0000}{\color[HTML]{FFFFFF} \ding{55}} & \cellcolor[HTML]{CB0000}{\color[HTML]{FFFFFF} \ding{55}} & \cellcolor[HTML]{009901}{\color[HTML]{FFFFFF} \ding{51}} \\ 
\hline
\multirow{2}{*}{Pattern} & \textbf{NLQ/A} & \textit{2017} & \cellcolor[HTML]{009901}{\color[HTML]{FFFFFF} \ding{51}} & \cellcolor[HTML]{009901}{\color[HTML]{FFFFFF} \ding{51}} & \cellcolor[HTML]{C0C0C0}{\color[HTML]{FFFFFF} \textbf{?}} & \cellcolor[HTML]{009901}{\color[HTML]{FFFFFF} \ding{51}} & \cellcolor[HTML]{009901}{\color[HTML]{FFFFFF} \ding{51}} & \cellcolor[HTML]{009901}{\color[HTML]{FFFFFF} \ding{51}} & \cellcolor[HTML]{009901}{\color[HTML]{FFFFFF} \ding{51}} & \cellcolor[HTML]{C0C0C0}{\color[HTML]{FFFFFF} \textbf{?}} & \cellcolor[HTML]{C0C0C0}{\color[HTML]{FFFFFF} \textbf{?}} & \cellcolor[HTML]{C0C0C0}{\color[HTML]{FFFFFF} \textbf{?}} & \cellcolor[HTML]{CB0000}{\color[HTML]{FFFFFF} \ding{55}} & \cellcolor[HTML]{009901}{\color[HTML]{FFFFFF} \ding{51}} \\
&\textbf{QuestIO} & \textit{2008} & \cellcolor[HTML]{009901}{\color[HTML]{FFFFFF} \ding{51}} & \cellcolor[HTML]{CB0000}{\color[HTML]{FFFFFF} \ding{55}} & \cellcolor[HTML]{CB0000}{\color[HTML]{FFFFFF} \ding{55}} & \cellcolor[HTML]{CB0000}{\color[HTML]{FFFFFF} \ding{55}} & \cellcolor[HTML]{009901}{\color[HTML]{FFFFFF} \ding{51}} & \cellcolor[HTML]{C0C0C0}{\color[HTML]{FFFFFF} \textbf{?}} & \cellcolor[HTML]{CB0000}{\color[HTML]{FFFFFF} \ding{55}} & \cellcolor[HTML]{C0C0C0}{\color[HTML]{FFFFFF} \textbf{?}} & \cellcolor[HTML]{CB0000}{\color[HTML]{FFFFFF} \ding{55}} & \cellcolor[HTML]{CB0000}{\color[HTML]{FFFFFF} \ding{55}} & \cellcolor[HTML]{CB0000}{\color[HTML]{FFFFFF} \ding{55}} & \cellcolor[HTML]{009901}{\color[HTML]{FFFFFF} \ding{51}} \\ 
\hline
\multirow{8}{*}{Parsing} & \textbf{ATHENA} & \textit{2016} & \cellcolor[HTML]{009901}{\color[HTML]{FFFFFF} \ding{51}} & \cellcolor[HTML]{009901}{\color[HTML]{FFFFFF} \ding{51}} & \cellcolor[HTML]{009901}{\color[HTML]{FFFFFF} \ding{51}} & \cellcolor[HTML]{009901}{\color[HTML]{FFFFFF} \ding{51}} & \cellcolor[HTML]{009901}{\color[HTML]{FFFFFF} \ding{51}} & \cellcolor[HTML]{009901}{\color[HTML]{FFFFFF} \ding{51}} & \cellcolor[HTML]{F8A102}{\color[HTML]{FFFFFF} \ding{115}} & \cellcolor[HTML]{CB0000}{\color[HTML]{FFFFFF} \ding{55}} & \cellcolor[HTML]{009901}{\color[HTML]{FFFFFF} \ding{51}} & \cellcolor[HTML]{CB0000}{\color[HTML]{FFFFFF} \ding{55}}  & \cellcolor[HTML]{009901}{\color[HTML]{FFFFFF} \ding{51}} & \cellcolor[HTML]{CB0000}{\color[HTML]{FFFFFF} \ding{55}}\\
&\textbf{Querix} & \textit{2006} & \cellcolor[HTML]{009901}{\color[HTML]{FFFFFF} \ding{51}} & \cellcolor[HTML]{C0C0C0}{\color[HTML]{FFFFFF} \textbf{?}} & \cellcolor[HTML]{C0C0C0}{\color[HTML]{FFFFFF} \textbf{?}} & \cellcolor[HTML]{009901}{\color[HTML]{FFFFFF} \ding{51}} & \cellcolor[HTML]{009901}{\color[HTML]{FFFFFF} \ding{51}} & \cellcolor[HTML]{C0C0C0}{\color[HTML]{FFFFFF} \textbf{?}} & \cellcolor[HTML]{C0C0C0}{\color[HTML]{FFFFFF} \textbf{?}} & \cellcolor[HTML]{C0C0C0}{\color[HTML]{FFFFFF} \textbf{?}} & \cellcolor[HTML]{C0C0C0}{\color[HTML]{FFFFFF} \textbf{?}} & \cellcolor[HTML]{C0C0C0}{\color[HTML]{FFFFFF} \textbf{?}} & \cellcolor[HTML]{CB0000}{\color[HTML]{FFFFFF} \ding{55}} & \cellcolor[HTML]{009901}{\color[HTML]{FFFFFF} \ding{51}} \\
&\textbf{FREyA} & \textit{2010} & \cellcolor[HTML]{009901}{\color[HTML]{FFFFFF} \ding{51}} & \cellcolor[HTML]{C0C0C0}{\color[HTML]{FFFFFF} \textbf{?}} & \cellcolor[HTML]{C0C0C0}{\color[HTML]{FFFFFF} \textbf{?}} & \cellcolor[HTML]{009901}{\color[HTML]{FFFFFF} \ding{51}} & \cellcolor[HTML]{C0C0C0}{\color[HTML]{FFFFFF} \textbf{?}} & \cellcolor[HTML]{C0C0C0}{\color[HTML]{FFFFFF} \textbf{?}} & \cellcolor[HTML]{C0C0C0}{\color[HTML]{FFFFFF} \textbf{?}} & \cellcolor[HTML]{CB0000}{\color[HTML]{FFFFFF} \ding{55}} & \cellcolor[HTML]{C0C0C0}{\color[HTML]{FFFFFF} \textbf{?}} & \cellcolor[HTML]{C0C0C0}{\color[HTML]{FFFFFF} \textbf{?}} & \cellcolor[HTML]{CB0000}{\color[HTML]{FFFFFF} \ding{55}} & \cellcolor[HTML]{009901}{\color[HTML]{FFFFFF} \ding{51}} \\ 
&\textbf{BELA} & \textit{2012} & \cellcolor[HTML]{009901}{\color[HTML]{FFFFFF} \ding{51}} & \cellcolor[HTML]{C0C0C0}{\color[HTML]{FFFFFF} \textbf{?}} & \cellcolor[HTML]{C0C0C0}{\color[HTML]{FFFFFF} \textbf{?}} & \cellcolor[HTML]{C0C0C0}{\color[HTML]{FFFFFF} \textbf{?}} & \cellcolor[HTML]{C0C0C0}{\color[HTML]{FFFFFF} \textbf{?}} & \cellcolor[HTML]{C0C0C0}{\color[HTML]{FFFFFF} \textbf{?}} & \cellcolor[HTML]{C0C0C0}{\color[HTML]{FFFFFF} \textbf{?}} & \cellcolor[HTML]{C0C0C0}{\color[HTML]{FFFFFF} \textbf{?}} & \cellcolor[HTML]{C0C0C0}{\color[HTML]{FFFFFF} \textbf{?}} & \cellcolor[HTML]{C0C0C0}{\color[HTML]{FFFFFF} \textbf{?}} & \cellcolor[HTML]{CB0000}{\color[HTML]{FFFFFF} \ding{55}} & \cellcolor[HTML]{009901}{\color[HTML]{FFFFFF} \ding{51}} \\
&\textbf{USI Answers} & \textit{2013} & \cellcolor[HTML]{009901}{\color[HTML]{FFFFFF} \ding{51}} & \cellcolor[HTML]{009901}{\color[HTML]{FFFFFF} \ding{51}} & \cellcolor[HTML]{009901}{\color[HTML]{FFFFFF} \ding{51}} & \cellcolor[HTML]{C0C0C0}{\color[HTML]{FFFFFF} \textbf{?}} & \cellcolor[HTML]{009901}{\color[HTML]{FFFFFF} \ding{51}} & \cellcolor[HTML]{009901}{\color[HTML]{FFFFFF} \ding{51}} & \cellcolor[HTML]{F8A102}{\color[HTML]{FFFFFF} \ding{115}} & \cellcolor[HTML]{009901}{\color[HTML]{FFFFFF} \ding{51}} & \cellcolor[HTML]{C0C0C0}{\color[HTML]{FFFFFF} \textbf{?}} & \cellcolor[HTML]{C0C0C0}{\color[HTML]{FFFFFF} \textbf{?}} & \cellcolor[HTML]{009901}{\color[HTML]{FFFFFF} \ding{51}} & \cellcolor[HTML]{009901}{\color[HTML]{FFFFFF} \ding{51}}\\ 
&\textbf{NaLIR (NaLIX)} & \textit{2014} & \cellcolor[HTML]{009901}{\color[HTML]{FFFFFF} \ding{51}} & \cellcolor[HTML]{009901}{\color[HTML]{FFFFFF} \ding{51}} & \cellcolor[HTML]{C0C0C0}{\color[HTML]{FFFFFF} \textbf{?}} & \cellcolor[HTML]{009901}{\color[HTML]{FFFFFF} \ding{51}} & \cellcolor[HTML]{009901}{\color[HTML]{FFFFFF} \ding{51}} & \cellcolor[HTML]{CB0000}{\color[HTML]{FFFFFF} \ding{55}} & \cellcolor[HTML]{009901}{\color[HTML]{FFFFFF} \ding{51}} & \cellcolor[HTML]{009901}{\color[HTML]{FFFFFF} \ding{51}} & \cellcolor[HTML]{009901}{\color[HTML]{FFFFFF} \ding{51}} & \cellcolor[HTML]{009901}{\color[HTML]{FFFFFF} \ding{51}}  & \cellcolor[HTML]{009901}{\color[HTML]{FFFFFF} \ding{51}} & \cellcolor[HTML]{CB0000}{\color[HTML]{FFFFFF} \ding{55}}\\
&\textbf{BioSmart} & \textit{2017} & \cellcolor[HTML]{009901}{\color[HTML]{FFFFFF} \ding{51}} & \cellcolor[HTML]{C0C0C0}{\color[HTML]{FFFFFF} \textbf{?}} & \cellcolor[HTML]{C0C0C0}{\color[HTML]{FFFFFF} \textbf{?}} & \cellcolor[HTML]{C0C0C0}{\color[HTML]{FFFFFF} \textbf{?}} & \cellcolor[HTML]{009901}{\color[HTML]{FFFFFF} \ding{51}} & \cellcolor[HTML]{009901}{\color[HTML]{FFFFFF} \ding{51}} & \cellcolor[HTML]{C0C0C0}{\color[HTML]{FFFFFF} \textbf{?}} & \cellcolor[HTML]{C0C0C0}{\color[HTML]{FFFFFF} \textbf{?}} & \cellcolor[HTML]{C0C0C0}{\color[HTML]{FFFFFF} \textbf{?}} & \cellcolor[HTML]{C0C0C0}{\color[HTML]{FFFFFF} \textbf{?}}  & \cellcolor[HTML]{009901}{\color[HTML]{FFFFFF} \ding{51}} & \cellcolor[HTML]{CB0000}{\color[HTML]{FFFFFF} \ding{55}}\\ 
\hline
\multirow{7}{*}{Grammar} & \textbf{TR Discover} & \textit{2015} & \cellcolor[HTML]{009901}{\color[HTML]{FFFFFF} \ding{51}} & \cellcolor[HTML]{C0C0C0}{\color[HTML]{FFFFFF} \textbf{?}} & \cellcolor[HTML]{C0C0C0}{\color[HTML]{FFFFFF} \textbf{?}} & \cellcolor[HTML]{CB0000}{\color[HTML]{FFFFFF} \ding{55}} & \cellcolor[HTML]{F8A102}{\color[HTML]{FFFFFF} \ding{115}} & \cellcolor[HTML]{C0C0C0}{\color[HTML]{FFFFFF} \textbf{?}} & \cellcolor[HTML]{CB0000}{\color[HTML]{FFFFFF} \ding{55}} & \cellcolor[HTML]{F8A102}{\color[HTML]{FFFFFF} \ding{115}} & \cellcolor[HTML]{C0C0C0}{\color[HTML]{FFFFFF} \textbf{?}} & \cellcolor[HTML]{C0C0C0}{\color[HTML]{FFFFFF} \textbf{?}} & \cellcolor[HTML]{009901}{\color[HTML]{FFFFFF} \ding{51}} & \cellcolor[HTML]{009901}{\color[HTML]{FFFFFF} \ding{51}}\\
&\textbf{Ginseng} & \textit{2005} & \cellcolor[HTML]{009901}{\color[HTML]{FFFFFF} \ding{51}} & \cellcolor[HTML]{C0C0C0}{\color[HTML]{FFFFFF} \textbf{?}} & \cellcolor[HTML]{C0C0C0}{\color[HTML]{FFFFFF} \textbf{?}} & \cellcolor[HTML]{C0C0C0}{\color[HTML]{FFFFFF} \textbf{?}} & \cellcolor[HTML]{009901}{\color[HTML]{FFFFFF} \ding{51}} & \cellcolor[HTML]{C0C0C0}{\color[HTML]{FFFFFF} \textbf{?}} & \cellcolor[HTML]{C0C0C0}{\color[HTML]{FFFFFF} \textbf{?}} & \cellcolor[HTML]{C0C0C0}{\color[HTML]{FFFFFF} \textbf{?}} & \cellcolor[HTML]{C0C0C0}{\color[HTML]{FFFFFF} \textbf{?}} & \cellcolor[HTML]{C0C0C0}{\color[HTML]{FFFFFF} \textbf{?}} & \cellcolor[HTML]{CB0000}{\color[HTML]{FFFFFF} \ding{55}} & \cellcolor[HTML]{009901}{\color[HTML]{FFFFFF} \ding{51}} \\
&\textbf{SQUALL} & \textit{2014} & \cellcolor[HTML]{F8A102}{\color[HTML]{FFFFFF} \ding{115}} & \cellcolor[HTML]{F8A102}{\color[HTML]{FFFFFF} \ding{115}} & \cellcolor[HTML]{F8A102}{\color[HTML]{FFFFFF} \ding{115}} & \cellcolor[HTML]{F8A102}{\color[HTML]{FFFFFF} \ding{115}} & \cellcolor[HTML]{F8A102}{\color[HTML]{FFFFFF} \ding{115}} & \cellcolor[HTML]{F8A102}{\color[HTML]{FFFFFF} \ding{115}} & \cellcolor[HTML]{F8A102}{\color[HTML]{FFFFFF} \ding{115}} & \cellcolor[HTML]{F8A102}{\color[HTML]{FFFFFF} \ding{115}} & \cellcolor[HTML]{F8A102}{\color[HTML]{FFFFFF} \ding{115}} & \cellcolor[HTML]{F8A102}{\color[HTML]{FFFFFF} \ding{115}} & \cellcolor[HTML]{CB0000}{\color[HTML]{FFFFFF} \ding{55}} & \cellcolor[HTML]{009901}{\color[HTML]{FFFFFF} \ding{51}} \\
&\textbf{MEANS} & \textit{2015} & \cellcolor[HTML]{009901}{\color[HTML]{FFFFFF} \ding{51}} & \cellcolor[HTML]{CB0000}{\color[HTML]{FFFFFF} \ding{55}} & \cellcolor[HTML]{C0C0C0}{\color[HTML]{FFFFFF} \textbf{?}} & \cellcolor[HTML]{CB0000}{\color[HTML]{FFFFFF} \ding{55}} & \cellcolor[HTML]{C0C0C0}{\color[HTML]{FFFFFF} \textbf{?}} & \cellcolor[HTML]{009901}{\color[HTML]{FFFFFF} \ding{51}} & \cellcolor[HTML]{CB0000}{\color[HTML]{FFFFFF} \ding{55}} & \cellcolor[HTML]{CB0000}{\color[HTML]{FFFFFF} \ding{55}} & \cellcolor[HTML]{CB0000}{\color[HTML]{FFFFFF} \ding{55}} & \cellcolor[HTML]{CB0000}{\color[HTML]{FFFFFF} \ding{55}} & \cellcolor[HTML]{CB0000}{\color[HTML]{FFFFFF} \ding{55}} & \cellcolor[HTML]{009901}{\color[HTML]{FFFFFF} \ding{51}} \\
&\textbf{AskNow} & \textit{2016} & \cellcolor[HTML]{009901}{\color[HTML]{FFFFFF} \ding{51}} & \cellcolor[HTML]{C0C0C0}{\color[HTML]{FFFFFF} \textbf{?}} & \cellcolor[HTML]{009901}{\color[HTML]{FFFFFF} \ding{51}} & \cellcolor[HTML]{C0C0C0}{\color[HTML]{FFFFFF} \textbf{?}} & \cellcolor[HTML]{009901}{\color[HTML]{FFFFFF} \ding{51}} & \cellcolor[HTML]{F8A102}{\color[HTML]{FFFFFF} \ding{115}} & \cellcolor[HTML]{C0C0C0}{\color[HTML]{FFFFFF} \textbf{?}} & \cellcolor[HTML]{C0C0C0}{\color[HTML]{FFFFFF} \textbf{?}} & \cellcolor[HTML]{009901}{\color[HTML]{FFFFFF} \ding{51}} & \cellcolor[HTML]{C0C0C0}{\color[HTML]{FFFFFF} \textbf{?}} & \cellcolor[HTML]{CB0000}{\color[HTML]{FFFFFF} \ding{55}} & \cellcolor[HTML]{009901}{\color[HTML]{FFFFFF} \ding{51}} \\
&\textbf{SPARKLIS} & \textit{2017} & \cellcolor[HTML]{009901}{\color[HTML]{FFFFFF} \ding{51}} & \cellcolor[HTML]{009901}{\color[HTML]{FFFFFF} \ding{51}} & \cellcolor[HTML]{009901}{\color[HTML]{FFFFFF} \ding{51}} & \cellcolor[HTML]{F8A102}{\color[HTML]{FFFFFF} \ding{115}} & \cellcolor[HTML]{009901}{\color[HTML]{FFFFFF} \ding{51}} & \cellcolor[HTML]{C0C0C0}{\color[HTML]{FFFFFF} \textbf{?}} & \cellcolor[HTML]{F8A102}{\color[HTML]{FFFFFF} \ding{115}} & \cellcolor[HTML]{009901}{\color[HTML]{FFFFFF} \ding{51}} & \cellcolor[HTML]{009901}{\color[HTML]{FFFFFF} \ding{51}} & \cellcolor[HTML]{009901}{\color[HTML]{FFFFFF} \ding{51}} & \cellcolor[HTML]{CB0000}{\color[HTML]{FFFFFF} \ding{55}} & \cellcolor[HTML]{009901}{\color[HTML]{FFFFFF} \ding{51}} \\
&\textbf{GFMed} & \textit{2017} & \cellcolor[HTML]{009901}{\color[HTML]{FFFFFF} \ding{51}} & \cellcolor[HTML]{CB0000}{\color[HTML]{FFFFFF} \ding{55}} & \cellcolor[HTML]{CB0000}{\color[HTML]{FFFFFF} \ding{55}} & \cellcolor[HTML]{009901}{\color[HTML]{FFFFFF} \ding{51}} & \cellcolor[HTML]{C0C0C0}{\color[HTML]{FFFFFF} \textbf{?}} & \cellcolor[HTML]{C0C0C0}{\color[HTML]{FFFFFF} \textbf{?}} & \cellcolor[HTML]{C0C0C0}{\color[HTML]{FFFFFF} \textbf{?}} & \cellcolor[HTML]{009901}{\color[HTML]{FFFFFF} \ding{51}} & \cellcolor[HTML]{C0C0C0}{\color[HTML]{FFFFFF} \textbf{?}} & \cellcolor[HTML]{C0C0C0}{\color[HTML]{FFFFFF} \textbf{?}} & \cellcolor[HTML]{CB0000}{\color[HTML]{FFFFFF} \ding{55}} & \cellcolor[HTML]{009901}{\color[HTML]{FFFFFF} \ding{51}} \\ 
\hline
\multirow{3}{*}{Commercial} & \textbf{Google} & \textit{} & \cellcolor[HTML]{009901}{\color[HTML]{FFFFFF} \ding{51}} & \cellcolor[HTML]{F8A102}{\color[HTML]{FFFFFF} \ding{115}} & \cellcolor[HTML]{CB0000}{\color[HTML]{FFFFFF} \ding{55}} & \cellcolor[HTML]{009901}{\color[HTML]{FFFFFF} \ding{51}} & \cellcolor[HTML]{009901}{\color[HTML]{FFFFFF} \ding{51}} & \cellcolor[HTML]{F8A102}{\color[HTML]{FFFFFF} \ding{115}} & \cellcolor[HTML]{F8A102}{\color[HTML]{FFFFFF} \ding{115}} & \cellcolor[HTML]{009901}{\color[HTML]{FFFFFF} \ding{51}} & \cellcolor[HTML]{009901}{\color[HTML]{FFFFFF} \ding{51}} & \cellcolor[HTML]{009901}{\color[HTML]{FFFFFF} \ding{51}} & \cellcolor[HTML]{C0C0C0}{\color[HTML]{FFFFFF} \textbf{?}} & \cellcolor[HTML]{C0C0C0}{\color[HTML]{FFFFFF} \textbf{?}} \\
& \textbf{Siri} & \textit{} & \cellcolor[HTML]{009901}{\color[HTML]{FFFFFF} \ding{51}} & \cellcolor[HTML]{CB0000}{\color[HTML]{FFFFFF} \ding{55}} & \cellcolor[HTML]{F8A102}{\color[HTML]{FFFFFF} \ding{115}} & \cellcolor[HTML]{CB0000}{\color[HTML]{FFFFFF} \ding{55}} & \cellcolor[HTML]{CB0000}{\color[HTML]{FFFFFF} \ding{55}} & \cellcolor[HTML]{CB0000}{\color[HTML]{FFFFFF} \ding{55}} & \cellcolor[HTML]{CB0000}{\color[HTML]{FFFFFF} \ding{55}} & \cellcolor[HTML]{CB0000}{\color[HTML]{FFFFFF} \ding{55}} & \cellcolor[HTML]{CB0000}{\color[HTML]{FFFFFF} \ding{55}} & \cellcolor[HTML]{CB0000}{\color[HTML]{FFFFFF} \ding{55}} & \cellcolor[HTML]{C0C0C0}{\color[HTML]{FFFFFF} \textbf{?}} & \cellcolor[HTML]{C0C0C0}{\color[HTML]{FFFFFF} \textbf{?}} \\
& \textbf{IMDb} & \textit{} & \cellcolor[HTML]{F8A102}{\color[HTML]{FFFFFF} \ding{115}} & \cellcolor[HTML]{009901}{\color[HTML]{FFFFFF} \ding{51}} & \cellcolor[HTML]{009901}{\color[HTML]{FFFFFF} \ding{51}} & \cellcolor[HTML]{CB0000}{\color[HTML]{FFFFFF} \ding{55}} & \cellcolor[HTML]{009901}{\color[HTML]{FFFFFF} \ding{51}} & \cellcolor[HTML]{CB0000}{\color[HTML]{FFFFFF} \ding{55}} & \cellcolor[HTML]{CB0000}{\color[HTML]{FFFFFF} \ding{55}} & \cellcolor[HTML]{CB0000}{\color[HTML]{FFFFFF} \ding{55}} & \cellcolor[HTML]{CB0000}{\color[HTML]{FFFFFF} \ding{55}} & \cellcolor[HTML]{CB0000}{\color[HTML]{FFFFFF} \ding{55}} & \cellcolor[HTML]{C0C0C0}{\color[HTML]{FFFFFF} \textbf{?}} & \cellcolor[HTML]{C0C0C0}{\color[HTML]{FFFFFF} \textbf{?}} 
\end{tabular}
\end{table*}

In general, as shown in Table \ref{tab:questions}, we can say that keyword-based NLIs are the least powerful and can only answer simple questions like string filters (\texttt{Q1}). This limitation is based on the approach of these keyword-based systems: they expect just keywords (which are mostly filters) and the systems identify relationships between them. Therefore, they do not expect any complexer questions like \texttt{Q4} or \texttt{Q7}.
Pattern-based NLIs are an extension of keyword-based systems in such a way that they have a dictionary with trigger words to answer more complex questions like aggregations (\texttt{Q7}). However, they cannot answer questions of higher difficulties, including subqueries (\texttt{Q9}/\texttt{Q10}). For example, the difficulty with questions including subqueries is to identify which part of the input question belongs to which subquery. Trigger words are not sufficient to identify the range of each subquery.

Parsing-based NLIs are able to answer these questions by using dependency or constituency trees (e.g., NaLIR). This helps to identify and group the input question in such a way that the different parts of the subqueries can be identified. Still, some of those systems struggle with the same problem as pattern-based NLIs: the systems are using trigger word dictionaries to identify certain types of questions like aggregations (e.g., ATHENA), which is not a general solution of the problem.

Grammar-based systems offer the possibility to guide the users during the formulation of their questions. Dynamically applying the rules during typing allows the systems to ensure that the input question is always translatable into formal language. The huge disadvantage of grammar-based systems is that they need handcrafted rules. There are NLIs which use a set of general rules and domain-specific rules (e.g., SQUALL). The general rules can be used for other domains and therefore increase the adaptability of the system. Other systems try to extract the rules directly from the ontology and thereby reduce the number of handcrafted rules (e.g., Ginseng).

We will now analyze how well the different systems can handle the ten sample questions. The first question is a basic filter question and can be solved by all NLIs as shown in Table \ref{tab:questions}. The last three questions are the most difficult ones and can only be answered by a few systems (completely by SQUALL, SPARKLIS and partially by others). Complex questions (e.g., aggregations) cannot be phrased with keywords only. Therefore, the more complicated the users questions are, the more they will phrase them in grammatically correct sentences. In contrast, simple questions (e.g., string filters like \texttt{Q1}) can be easily asked with keywords. Both, \cite{USIAnswers2013} (USI Answer) and \cite{NLmaps2016} (NLmaps) are describing this phenomenon and that the users prefer to ask questions with keywords if possible. They adapted their systems so that they can handle different forms of user input. Because of that, \cite{USIAnswers2013} (USI Answer) point out that parse trees should only be used with caution. This is similar to the approach of \cite{NLQ/A2017} (NLQ/A), who remark that NLP technologies are not worth the risk, because wrong interpretations in the processing leads to errors. \cite{BELA2012} (BELA) propose a new approach of applying certain processing steps only if the question cannot be answered by using simpler mechanisms. This approach can be used to answer questions formulated as keywords or as complete sentences. Nevertheless, parse trees are useful to identify subqueries, but only in grammatically correct sentences (e.g., NaLIR). The identification of possible subqueries is necessary to answer questions like \texttt{Q9} and \texttt{Q10}.

Based on the sample questions, SQUALL, SPARKLIS, NaLIR and ATHENA are the systems that perform best (i.e., they can handle most of the question types). However, these systems still have some drawbacks. SQUALL requires that the user has knowledge of the RDF vocabulary. For example, if the users ask about all movies starring Brad Pitt, the question needs to be similar to `\textit{All \texttt{movie-s} starring \texttt{Brad\_Pitt}}'. \linebreak SPARKLIS can answer all questions (if concepts are defined) but is based on a strict user interface where the users have to `click their questions together' and cannot `write freely'. In contrast to those two systems, NaLIR and ATHENA are systems where the user can write without restrictions during the process of phrasing the questions. However, NaLIR cannot handle concepts. Finally, ATHENA solves aggregations with trigger words which the users need to know. Moreover, ATHENA cannot solve multiple subqueries.

\subsection{Evaluation of commercial systems}
\label{subsec:commercial}
Increasingly, natural language interfaces are deployed for commercial usage. We decided to ask the ten sample questions to three commercial systems: Google\footnote{\url{https://www.google.com/}}, Siri\footnote{\url{https://www.apple.com/siri/}} and Internet Movie Database (IMDb)\footnote{\url{https://www.imdb.com}}.

Based on the sample questions, Google is the best system in this category. An answer is assumed to be correct, if Google displays the answer in the featured snippet at the top of the results or in the knowledge panel on the side. For example, \texttt{Q1} is answered in the featured snippet. In contrast, for question \texttt{Q2} the featured snippet on top shows the top 250 drama movies, but the first result site contains the correct answer.

Siri can only answer simple \texttt{select} and \texttt{filter} questions and has some troubles to handle the year without a specific date in \texttt{Q3}. What is different to other systems is that if it cannot answer a question, Siri gives feedback to the user and tries to explain which type of questions can be answered. For most of the sample questions, we got the answer ``\textit{Sorry, I can't search what something is about. But I can search by title, actors or directors and categories like horror or action.}''

IMDb is a keyword-based system with the option of advanced searches\footnote{E.g. advanced title search \url{https://www.imdb.com/search/title}} which are form-based. This system is not able to provide precise answers to \texttt{select} questions like \texttt{Q1} about the director of a given movie. However, the users can find the correct results by browsing the movie page.

\section{Machine Learning Approaches for NLIs}
\label{sec:new_developments}

In current research, more and more systems include machine learning (ML) in their translation process (e.g., MEANS) or ranking (e.g., Aqqu). 

KBQA \citep{KBQA2017} learns templates as a kind of question representation. It supports binary factoid questions as well as complex questions which are composed of binary factoid questions. OQA \citep{OQA2014} approaches to leverage both curated and extracted KBs, by mining millions of rules from an unlabeled question corpus and across multiple KBs. 

AMUSE \citep{AMUSE2017} uses ML to determine the most probable meaning of a question and can handle different languages at the same time. Xser \citep{Xser2014} divides the translation task into a KB-independent and a KB-related task. In the KB-independent task, they are developing a \textit{Directed Acyclic Graph} parser to capture the structure of the query intentions and trained it on human labeled data. 

A new promising avenue of research is to use deep learning techniques as the foundation for NLIDBs. The basic idea is to formulate the translation of natural language (NL) to SQL as an end-to-end machine translation problem (\cite{sutskever2014sequence, dong2016language, jia2016data}). The approach is often called neural semantic parsing (\cite{wang2015building}). In other words, translating from NL to SQL can be formulated as a supervised machine learning problem on pairs of natural language and SQL queries. In particular, machine translation can be modeled as a sequence-to-sequence problem where the input sequence is represented by the words (tokens) of the NL and the output sequence by the tokens of SQL. The main goal is given an input sequence of tokens, predict the output sequence based on observed patterns in the past.  

The main advantage of machine learning based approaches over traditional NLIDBs is that they support a richer linguistic variability in query expressions and thus users can formulate queries with greater flexibility. However, one of the major challenges of supervised machine learning approaches is that they require a large training data set in order to achieve good accuracy on the translation task.

The most commonly used approach for sequence-to-sequence modeling is based on recurrent neural networks (RNNs, \cite{elman1990finding}) with an input encoder and an output decoder. The encoder and decoder are implemented as bi-directional LSTMs (Long Short Term Memory) by \cite{hochreiter1997long}. However, before an NL can be encoded, the tokens need to be represented as a vector that in turn can be processed by the neural network. A widely used approach is to use word embeddings where the vocabulary of the NL is mapped to a high-dimensional vector (\cite{pennington2014glove}). In order to improve the accuracy of the translation, attention models are often applied (\cite{luong2015effective}). 

One of the currently most advanced neural machine translation systems was introduced by \cite{iyer2017learning}. Their approach uses an encoder-decoder model with global attention similar to \cite{luong2015effective} and a bi-directional LSTM network to encode the input tokens. In order to improve the translation accuracy from NL to SQL compared to traditional machine translation approaches, the database schema is taken into account. Moreover, external paraphrases are used to increase the linguistic variance of the training data. In the first step of training the neural machine translation system, the process of generating training data is bootstrapped by manually-handcrafted templates for the NL and SQL query pairs. In the next phase, the training set is extended by adding linguistic variations of the input questions and parts to the query are replaced with synonyms or  paraphrases of the query. The advantage of this approach is that it is query language independent and could in principle also be used to translate from NL to SPARQL. However, the disadvantage is that a large, manually handcrafted a training set is necessary. 

\cite{zhong2017seq2sql} introduce a system called Seq2SQL. Their approach uses a deep neural network architecture with reinforcement learning to translate from NL to SQL. The authors released WikiSQL - a new data set based on Wikipedia consisting of 24,241 tables and 80,654 hand-annotated NL-SQL-pairs. However, their approach was only demonstrated to work on simple single-table queries without joins. SQLNet by \cite{xu2017sqlnet} uses a more traditional machine translation approach without reinforcement learning. However, even though SQLNet shows better accuracy than Seq2SQL, the experiments are also based on the WikiSQL data set and it is not clear how this approach would handle join queries against more realistic database settings with multiple tables. Finally, \cite{yavuz2018takes} show further improvements over SQLNet by incorporating both the information about the database schema as well as the base data. The paper in particular focuses on the generation of WHERE-clauses, which the authors identified as major problem of the relative low accuracy of SQL queries generated by Seq2SQL and SQLNet.

DBPal by \cite{basik2018dbpal} overcomes shortcomings of manually labeling large training data sets by synthetically generating a training set that only requires minimal annotations in the database. Similar to \cite{iyer2017learning}, DBPal uses the database schema and query templates to describe NL/SQL-pairs. Moreover,  inspired by \cite{wang2015building}, DBPal augments an initial set of NL queries using a paraphrasing approach based on a paraphrase dictionary. The results show that on a single-table data set DPal performs better than the semantic parsing approach of \cite{iyer2017learning}. However, for a multi-table data set with join queries, DBPal performs worse. The reason is that the limited training set does not seem to generalize well for join-queries. The authors attributed the good performance of the system introduced by \cite{iyer2017learning} to overfitting.

\cite{soru2017sparql} use neural machine translation approach similar to \cite{iyer2017learning}. However, the major difference is that they translate natural language to SPARQL rather than to SQL. Moreover, they do not apply an attention mechanism. In a subsequent white paper \cite{hartmann2018sparql} present an approach to automatically generate a large set of training data consisting of 894,499 training examples based on set of 5000 natural language queries. The approach is very similar to the approach used by DBPal.
 
In general, these new approaches show promising results, but they have either only been demonstrated to work for single-table data sets or require large amounts training data. Hence, the practical usage in realistic database settings still needs to be shown.

Another interesting new trend is in the area of conversational systems such as \cite{williams2015fast,john2017ava} that often apply neural machine translation techniques. However, a detailed discussion on these systems is beyond the scope of this paper.


\section{Conclusions}
\label{sec:conclusion}

In this paper, we summarized 24 recently developed natural language interfaces for databases. Each of them was evaluated using ten sample questions to show their strengths and weaknesses. Based on this evaluation, we identified the following lessons learned that are relevant for the development of NLIs.

\textbf{Lesson 1 - Use distinct mechanisms for handling simple versus complex questions:}
Users like to pose questions differently depending on the complexity of the question. Simple questions will often be asked with keywords, while complex questions are posed in grammatically correct sentences. For example, if users search for information about Brad Pitt, it is more likely that the users ask `\textit{Brad Pitt}' as an input question and not `\textit{Give me information about Brad Pitt}.'. This lesson is highlighted in the paper by \cite{USIAnswers2013} (USI Answer) and related to the finding that casual users are more at ease at posing complex questions in natural language than in other formats \citep{Kaufmann2010}. Generally, pattern-based systems are solving this problem, but are often limited in the understanding of complex questions (i.e., subqueries). \cite{BELA2012} (BELA) propose a layered system which could also be used to solve this problem of distinction. Grammar-based systems could be taught to support such non natural language question patterns when included in the grammar. To the best of our knowledge, none of the systems we looked at supports this feature.

\textbf{Lesson 2 - Identifying subqueries still is a significant challenge for NLIs:}
The identification of subqueries seems to one of the most difficult problems for NLIs. The sample input questions \texttt{Q9} and \texttt{Q10} are two examples for questions composed of one and multiple subqueries, respectively. The most common NLP technology that is able to solve this problem is a parse tree. This can either be a general dependency or constituency parse tree provided, for example, by the Stanford Parser (e.g., NaLIR), or a parse tree self-learned with the rules of a grammar-based NLI (e.g., SQUALL). An alternative is the use of templates (e.g., AskNow). \cite{NaLIR2014} (NaLIR) mention that the identification alone is not enough: after the identification of the subqueries, the necessary information needs to be propagated to each part of the subquery.

\textbf{Lesson 3 -  Optimize the number of user interactions:}
Natural language is ambiguous and even in a human to human conversation ambiguities occur, which are then clarified with questions. The same applies to NLIs: when an ambiguity occurs, the system needs to clarify with the user. This interaction should be optimized in such a way that the number of needed user interactions is minimized. To address this issue, \cite{QUICK2009} (QUICK) and \cite{NLQ/A2017} (NLQ/A) developed a minimization algorithm. The idea is to identify the ambiguity which has the most impact on the other ambiguities and clarify it first.

\textbf{Lesson 4 - Use grammar for guidance:}
The biggest advantage of grammar-based NLIs is that they can use their grammar rules to guide the users while they are typing their questions. This improves the interaction between system and users in two ways: first, the system will understand each question the users ask, second, the users will learn how certain questions have to be asked to receive a good result. For the second part, \cite{NaLIX2005} (NaLIX) propose a solution by using a question historization and templates. This also helps the users understand what type of questions the system can understand and how the users need to ask.

\textbf{Lesson 5 - Use hybrid approach of traditional NLI systems and neural machine translation}

New NLIs based on neural machine translation show promising results. However, the practical usage in realistic database settings still needs to be shown. Using a hybrid approach of traditional NLIs that are enhanced by neural machine translation might be a good approach for the future. Traditional approaches would guarantee better accuracy while neural machine translation approaches would increase the robustness to language variability. 

In summary, our evaluation of various systems against ten sample questions shows that NLIs have made significant progress over the last decade. However, our lessons also show that more research is required to increase the expressive power of NLIs. This paper gives a guide for researchers and practitioners highlighting the strengths and weaknesses of current approaches as well as helps them design and implement NLIs that are not only of theoretical value but have impact in industry.

\begin{acknowledgements}
The work was supported by Hasler Foundation under grant number 17040.
\end{acknowledgements}

\bibliographystyle{spbasic}      
\bibliography{bibtex.bib}   

\begin{appendices}
\section{SQL Representation}

\noindent\texttt{Q1:} 
\lstset{language=SQL,basicstyle=\ttfamily,} 
\begin{lstlisting}
SELECT DISTINCT p.*
FROM movie m, person p, directing d
WHERE m.id = d.movieId
  AND person.id = d.directorId
  AND m.title = 'Inglourious Basterds'
\end{lstlisting}

\noindent\texttt{Q2:} 
\lstset{language=SQL,basicstyle=\ttfamily,} 
\begin{lstlisting}
SELECT * FROM movie m WHERE m.rating > 9
\end{lstlisting}

\noindent\texttt{Q3:} 
\lstset{language=SQL,basicstyle=\ttfamily,} 
\begin{lstlisting}
SELECT DISTINCT m.*
FROM movie m, starring s, person p
WHERE m.id = s.movieId 
  AND s.actorId = p.id
  AND p.name = 'Brad Pitt'
  AND m.releaseDate >= '2000-01-01' 
  AND m.releaseDate <= '2010-12-31'
\end{lstlisting}

\noindent\texttt{Q4:} 
\lstset{language=SQL,basicstyle=\ttfamily,} 
\begin{lstlisting}
SELECT DISTINCT m.*
FROM movie m, gross g
WHERE m.id = g.movieId 
  AND g.gross = (
    SELECT MAX(gross) FROM gross
  )
\end{lstlisting}

\noindent\texttt{Q5:} 
\lstset{language=SQL,basicstyle=\ttfamily,} 
\begin{lstlisting}
SELECT DISTINCT m.*
FROM movie m, genre g
WHERE m.id = g.movieId
    AND (
      g.genre = 'drama' 
      OR g.genre = 'comedy'
    )
\end{lstlisting}

\noindent\texttt{Q6:} 
\lstset{language=SQL,basicstyle=\ttfamily,} 
\begin{lstlisting}
SELECT * FROM movie m WHERE m.rating >= 8
\end{lstlisting}

\noindent\texttt{Q7:} 
\lstset{language=SQL,basicstyle=\ttfamily,} 
\begin{lstlisting}
SELECT DISTINCT m.*
FROM movie m, genre g, (
    SELECT g.genre, max(m.rating) as maxRating 
    FROM movie m, genre g 
    WHERE m.id = g.moveId 
    GROUP BY g.genre
  ) as maxMovie
WHERE m.id = g.movieId 
  AND m.rating = maxMovie.maxRating
  AND g.genre = maxMovie.genre
\end{lstlisting}

\noindent\texttt{Q8:} 
\lstset{language=SQL,basicstyle=\ttfamily,} 
\begin{lstlisting}
SELECT DISTINCT m.*
FROM movie m, genre g
WHERE m.id = g.movieId
    AND originalLang != 'jp'
    AND g.genre = 'horror'
\end{lstlisting}

\noindent\texttt{Q9:} 
\lstset{language=SQL,basicstyle=\ttfamily,} 
\begin{lstlisting}
SELECT *
FROM movie m1
WHERE m1.rating > (
  SELECT m2.rating 
  FROM movie m2
  WHERE m2.title = 'Sin City'
)
\end{lstlisting}

\noindent\texttt{Q10:} 
\lstset{language=SQL,basicstyle=\ttfamily,} 
\begin{lstlisting}
SELECT DISTINCT m1.*
FROM genre g1, movie m1
WHERE m1.id = g1.movieId
  AND NOT EXISTS (
    SELECT ''
    FROM movie m2
    WHERE m2.title = 'Sin City'
      AND NOT EXISTS (
        SELECT  ''
        FROM genre g2
        WHERE g2.movieId = m1.id 
          AND g2.id = g1.id
      )
  )
\end{lstlisting}

\end{appendices}

\end{document}